\documentclass[11pt, a4paper]{article}
\usepackage[english]{babel}
\usepackage[latin1]{inputenc}
\usepackage[tbtags]{amsmath}
\usepackage{amsfonts,euscript,amssymb,stmaryrd,braket}
\usepackage{graphics,tikz}
\usetikzlibrary{arrows,decorations.markings,patterns}
\usepackage{slashed}
\definecolor{hyperref}{RGB}{026,028,185}
\usepackage[bookmarks=true,colorlinks=true,linkcolor=hyperref,citecolor=hyperref,urlcolor=hyperref,bookmarksnumbered]{hyperref}
\usepackage{cite}
\numberwithin{equation}{section}

\newcommand{\be}{\begin{equation}}
\newcommand{\ee}{\end{equation}}

\newcommand{\pa}{\partial}

\newcommand{\Tr}{\operatorname{Tr}}

\newcommand{\g}{\gamma}
\renewcommand{\d}{\delta}

\newcommand{\e}{\epsilon}

\newcommand{\p}{\pi}

\newcommand{\s}{\sigma}

\renewcommand{\t}{\tau}

\newcommand{\vf}{\varphi}

\newcommand{\cp}{\mathbb{CP}}

\newcommand{\adscp}{$AdS_4\times \mathbb{CP}^3$ }
 \newcommand{\abjm}{{\rm ABJM}}
  \newcommand{\YM}{{\rm YM}}

\begin{document}
%%%%%%%%%%%%%%%%%%%%%%%%%%%%%%%%%

%%%%%%%%%%%%%%%%%%%%%%%%%%%%%%%%%
\thispagestyle{empty}

\vspace{ -3cm} \thispagestyle{empty} \vspace{-1cm}
\begin{flushright}
\footnotesize
HU-EP-14/30
\end{flushright}

\begingroup\centering
{\Large\bfseries
Two-loop cusp anomaly in ABJM at strong coupling
\par}
\vspace{7mm}

\begingroup
Lorenzo~Bianchi$^{a,1}$, Marco S. Bianchi$^{a,1}$, Alexis Br\`es$^{a,b,2}$,\\ Valentina Forini$^{a,1}$ and Edoardo Vescovi$^{a,1}$\\
\endgroup
\vspace{8mm}
\begingroup\small
\emph{$^{a}$ Institut f\"ur Physik,
Humboldt-Universit\"at
zu Berlin\\ Newtonstra{\ss}e 15, 12489 Berlin, Germany}\\
\emph{$^{b}$ Master ICFP, D\'epartement de Physique, \'Ecole Normale Sup\'erieure\\
24 rue Lhomond, 75231 Paris, France.}\\
\endgroup

\vspace{0.3cm}
\begingroup\small
$^{1}$ {\tt $\{$lorenzo.bianchi, marco.bianchi, valentina.forini, edoardo.vescovi$\}$@\,physik.hu-berlin.de}\\
$^{2}$ {\tt alexis.bres@\,ens.fr}
\endgroup
\vspace{1.0cm}

\textbf{Abstract}\vspace{5mm}\par
\begin{minipage}{14.7cm}
We compute the null cusp anomalous dimension of ABJM theory at strong coupling up to
two-loop order. This is done by evaluating  corrections to the corresponding superstring 
partition function, weighted by the \adscp action in AdS  light-cone gauge.  
%All occurring one- and two-loop UV divergencies are shown to cancel, 
%with a  mechanism that closely resembles its $AdS_5\times S^5$ counterpart. 
We compare our  result, where we use an anomalous shift in the $AdS_4$ radius, 
with the  cusp anomaly of $\mathcal{N}=4$ SYM, 
%in the way predicted by the $AdS_4/$CFT$_3$ Bethe ansatz. From this we can 
and extract the two-loop contribution to the non-trivial integrable coupling $h(\lambda)$ 
of ABJM theory. It coincides with the strong coupling expansion of
 the exact expression for $ h(\lambda)$ recently conjectured by Gromov and Sizov.  Our work provides 
 thus a 
 non-trivial perturbative check for the latter, as well as evidence 
 for two-loop UV-finiteness and quantum integrability of  
 the  Type IIA \adscp superstring in this gauge.
 
\end{minipage}\par
\endgroup
%%%%%%%%%%%%%%%%%%%%%%%%%%%%%%%%%

\newpage

%%%%%%%%%%%%%%%%%%%%%%%%%%%%%%%%%
\tableofcontents
%%%%%%%%%%%%%%%%%%%%%%%%%%%%%%%%%

\newpage

%%%%%%%%%%%%%%%%%%%%%%%%%%%%%%%%%
\section{Overview and results}\label{intro}
%%%%%%%%%%%%%%%%%%%%%%%%%%%%%%%%%

A  powerful attribute that the planar AdS$_4/$CFT$_3$ system~\cite{Aharony:2008ug} shares with its 
higher-dimensional version, planar AdS$_5/$CFT$_4$~\cite{Maldacena:1997re}, is the conjectured 
integrability~\cite{Minahan:2008hf,Gromov:2008qe,Klose:2010ki, Cavaglia:2014exa}
 of the gauge and (free) string theory model that define it --
respectively $\mathcal{N}=6$ super  Chern-Simons-matter (ABJM) theory in $d=3$
and Type IIA superstrings in a \adscp background with two- and four-form RR fluxes.
The explicit realization of the integrable structure is however non-trivial, due to significative 
peculiarites of this case. A first one is the absence of maximal supersymmetry 
in the \adscp background. This  makes the construction of the corresponding superstring 
 action difficult, in particular with  issues on  the $\kappa$-symmetry gauge-fixing suitable to 
 describe strings moving only in $AdS$, the latter being a relevant setting for  the studies of 
quantum integrability~\cite{Maldacena:1998im,Gubser:2002tv,Kruczenski:2002fb}. 
%In general, the more complicated structure of the $\cp^3$ 
%background results in more involved structures appearing 
%VF: cite

A second, crucial, peculiarity of the AdS$_4/$CFT$_3$ system is that all 
integrability-based calculations are  given in terms of a non-trivial, 
interpolating  function of the 't Hooft coupling $h(\lambda)$, 
 appearing in the ABJM magnon dispersion relation~\footnote{In the $\mathcal{N}=4$ SYM case the 
 relation of $h(\lambda_\YM)$ with the coupling  is trivial at all orders, $h(\lambda_\YM)=\sqrt{\lambda_\YM}/(4\pi)$, as
 shown in~\cite{ Correa:2012at, Correa:2012hh,Gromov:2012eu} by evaluating the so-called   ``Brehmstrahlung function'' 
both via an extrapolation on  results of supersymmetric localization and via integrability. 
See also discussions in~\cite{Hofman:2006xt,Klose:2007rz ,Berenstein:2009qd}.}
 \be
 \epsilon=\frac{1}{2}\sqrt{1+16\, h^2(\lambda)\sin^2\frac{p}{2}}\,.
 \ee
 Its knowledge is decisive to grant the conjectured integrability of ABJM theory a full predictive power.
 
 The first few orders of its weak coupling expansion were computed 
 in~\cite{Gaiotto:2008cg,Grignani:2008is,Nishioka:2008gz}
and in~\cite{Minahan:2009aq,Minahan:2009wg,Leoni:2010tb}.
%VF clarify references (and briefly method) 
At strong coupling, one way to obtain information on $h(\lambda)$ 
is to evaluate in string theory 
the universal scaling function~\footnote{Scaling function and cusp anomaly 
appear often as synonyms in the literature. 
At weak coupling and in the $\mathcal{N}=4$ case  the scaling function $f(\lambda_\YM)$, multiplying the $\log S$ in the large spin anomalous dimensions of twist-two 
operators,  equals twice the cusp anomalous dimension  $\Gamma_{\rm cusp}$ 
of light-like Wilson loops~\cite{Belitsky:2006en}. The same has been seen at strong coupling 
in~\cite{Kruczenski:2002fb, Kruczenski:2007cy}.} 
for the ABJM theory $f_{\abjm}(\lambda)$, and then 
compare the result obtained with the asymptotic Bethe ansatz prediction 
of~\cite{Gromov:2008qe}. The latter is based on the equivalence of the BES~\cite{Beisert:2006ez} equations for the $\mathcal{N}=4$ 
case and the ABJM case and reads
%~\footnote{The prediction is based on the equivalence of the BES equations for the $\mathcal{N}=4$ 
%case and the ABJM case, provided the replacement of the coupling constant as in \eqref{cusp_pred}.} 
\be\label{cusp_pred}
f_{\abjm}(\lambda)=\left.\frac{1}{2}\, f_{\mathcal{N}=4}(\lambda_\YM)\,\right|_{\frac{\sqrt{\lambda_\YM}}{4\pi}
\rightarrow h(\lambda)}\,,
\ee
which implies
\be\label{cusp_pred_2}
f_{\abjm}(\lambda)=2h(\lambda)-\frac{3\log 
2}{2\pi}-\frac{K}{8\pi^2}\frac{1}{h(\lambda)}+\cdots~,\\
\ee
where $f_{\mathcal{N}=4}(\lambda_\YM)$ is the cusp anomaly of $\mathcal{N}=4$ 
SYM  and $K$ is the Catalan constant. The leading strong coupling value for $f(\lambda)$ 
has been given already in~\cite{Aharony:2008ug} and reads 
$f(\lambda\gg1)=\sqrt{2\lambda}$, from which via \eqref{cusp_pred_2} one gets 
$h(\lambda\gg1)=\sqrt{\lambda/2}$. At one loop in sigma-model perturbation theory, 
the scaling function has been evaluated in~\cite{McLoughlin:2008ms,Alday:2008ut,Krishnan:2008zs,McLoughlin:2008he,Gromov:2008fy,
 Astolfi:2008ji,Bandres:2009kw, Abbott:2010yb,Abbott:2011xp,Astolfi:2011ju,Astolfi:2011bg,LopezArcos:2012gb,Forini:2012bb} 
via the energy of closed spinning strings  in the large spin limit or similar means, providing a first subleading
 correction $-\log2/(2\pi)$ to $h(\lambda)$ on which some debate existed~\cite{Shenderovich:2008bs}. In these calculations no 
issues were encountered in the action to use, as at one-loop only the quadratic 
part of the fermion Lagrangian is  necessary, with a structure which is well-known in terms of the 
  type IIA covariant derivative restricted by the background RR fluxes~\footnote{Alternatively, 
 one could still use the coset action of~\cite{Stefanski:2008ik, Arutyunov:2008if} - which is not suitable
 when strings move confined in AdS~\cite{Arutyunov:2008if, Gomis:2008jt} - starting with a classical  
 solution spinning both in $AdS_4$ with spin $S$ and in $\cp^3$ with spin $J$, 
 and taking on the resulting expression for the one-loop energy a smooth $J\to 0$ 
 limit~\cite{Alday:2008ut}.}.

We extend here the evaluation of the ABJM cusp anomaly  to the two-loop order in sigma-model 
perturbation theory using the open string 
approach~\cite{Kruczenski:2002fb,Kruczenski:2007cy} 
(in Type IIA), namely expanding the string 
partition function for the Euclidean surface ending on a null cusp at the boundary of $AdS_4$, as done in 
the $AdS_5\times S^5$ setting in~\cite{Giombi:2009gd}. 
As the classical string lies solely in $AdS_4$ and higher-order fermions are needed 
we must first face the problem, mentioned above, 
of using the correct  superstring action.  The  coset $OSp(6|4)/\left( U(3)\times SO(1, 3)\right)$ sigma-model formulation of it~\cite{Arutyunov:2008if, Stefanski:2008ik} is built following  
 the lines of (flat space and) type IIB 
 superstrings~\cite{Metsaev:1998it}, and  exhibits classical integrability. 
%{\bf tobe be completed, reduced AdS4-model was however shown to be classically integrable to all 
%orders in the fermions}.  
%~\footnote{See also~\cite{Astolfi:2009qh} on the discussion  on a  kappa-symmetry  gauge-fixing  
%consistent with the light-cone  gauge in the near plane- wave limit of \adscp.}
%As the problem with the supercoset action can be interpreted as a 
%exhibited in a particular choice $\kappa$-symmetry gauge-fixed action of  IIA GS string, 
%a way out is to start from it [uvarov grassi zarembo?]
 %To describe this dynamical sector one 
%should start  this happens due to  way out is to  start 
It  can be interpreted as a partially gauge-fixed type 
IIA Green Schwarz action, where the $\kappa$-symmetry gauge-fixing sets to zero eight fermionic 
modes corresponding to the eight broken supersymmetries.  
 However, as first argued in~\cite{Arutyunov:2008if} and 
 later clarified in~\cite{Gomis:2008jt},   it is not suitable 
 to describe the dynamics of a string lying solely in the $AdS_4$ part~\footnote{The same is true 
 when the string forms a worldsheet instanton by wrapping a $\cp^1$ cycle in $\cp^3$~\cite{Cagnazzo:2009zh}.} 
 of the   \adscp superspace, in that in this case four of the eight modes set to zero 
 are in fact dynamical fermionic degrees of freedom of the superstring.  Any action willing to capture 
 the semiclassical dynamics on these 
classical string configurations should contain these physical fermions, and should therefore be found 
via another, sensible $\kappa$-symmetry 
 gauge-fixing of the full action. This has been done 
 in~\cite{Uvarov:2009hf,Uvarov:2009nk}~\footnote{See also~\cite{Grassi:2009yj}.}
 %VF mention properly Gomis!
    %VF mention also Sorokin Wulff in ``Evidence"
    ,  starting  from the $D=11$ membrane action~\cite{deWit:1998yu} based on the supercoset 
    $OSp(8|4)/\left(SO(7)\times SO(1, 3)\right)$, 
 performing double dimensional reduction and choosing
 a $\kappa$-symmetry light-cone gauge for which both light-like directions lie in $AdS_4$.
 The output is an action, at most  quartic in the fermions, which is the  \adscp counterpart of the 
 gauge-fixed action of~\cite{Metsaev:2000yf,Metsaev:2000yu}. As the latter was
 efficiently used in~\cite{Giombi:2009gd} to evaluate the strong coupling corrections to the 
 $\mathcal{N}=4$ SYM cusp 
 anomaly up to two-loop order, the analysis of~\cite{Uvarov:2009hf,Uvarov:2009nk} is  
 the natural setup where 
 to perform our calculation.

Any known classical string solution  found  in $AdS_5$, which can be embedded within an $AdS_4$ 
subspace, is immediately a solution for this 
theory~\cite{Aharony:2008ug}. Therefore we start using the null cusp solution of~\cite{Kruczenski:2002fb, Giombi:2009gd} 
in the \adscp action of~\cite{Uvarov:2009hf, Uvarov:2009nk} and proceed evaluating 
corrections to the string path integral on it. 
These quantum string  corrections are 
in general non-trivial to calculate, in connection with issues of potential UV divergences and the lack of 
manifest power-counting renormalizability of the string action when expanded around a particular 
background (see discussion in~\cite{Roiban:2007jf,Giombi:2009gd,Giombi:2010fa, Giombi:2010zi})\footnote{In the evaluation of the 
worldsheet S-matrix starting from  the  light-cone gauge fixed $AdS_5\times S^5$ 
GS superstring action, non-cancellation of UV divergences has been observed already beyond the 
tree-level order  (see discussion in~\cite{Engelund:2013fja}). These issues, 
non present in alternative perturbative methods based on 
 unitarity cuts~\cite{Bianchi:2013nra,Engelund:2013fja,Bianchi:2014rfa},
 are still calling for an explanation. }, but have the additional important role 
 of establishing the quantum consistency of the proposed string actions. This is 
 a further motivation  for the study  at the quantum level of the action 
 proposed in~\cite{Uvarov:2009hf}, where the more complicated structure of the 
$\cp^3$ background translates in a considerably  more involved expression 
with respect to~\cite{Metsaev:2000yf,Metsaev:2000yu}. About the integrability 
of this string non-coset model, the standard analysis of~\cite{Bena:2003wd} - 
which applies to the action of~\cite{Arutyunov:2008if} - is not possible here. 
The  classical integrability of strings generically moving in 
the full \adscp superspace has been however shown by constructing
a Lax connection with zero curvature  up to  quadratic order in the 
fermions~\cite{Sorokin:2010wn}~\footnote{A study of classical 
integrability (prior to gauge-fixing) for general motion of the string in several backgrounds of 
interest for the AdS/CFT correspondence is in~\cite{Wulff:2014kja}.}.

Similarly to the $AdS_5\times S^5$ case, the $AdS$ light-cone approach  to the evaluation of 
the cusp anomaly turns out to be extremely efficient.  The background solution is ``homogeneous'', 
%-  i.e. the derivatives of the background fields are $(\tau, \sigma)$-independent - 
namely the  fluctuation Lagrangian turns out to have only constant coefficients. This makes immediate the study 
of the fluctuation  spectrum and highly simplifies the semiclassical analysis at higher orders~\footnote{
The evaluation of perturbative (sigma-model) string corrections for non-homogenous solutions  is currently 
limited to one-loop order, as in these cases in the fluctuation spectrum  (and thus in the propagator) 
non-trivial special elliptic functions
appear~\cite{Beccaria:2010ry,Forini:2010ek,
Drukker:2011za, Forini:2012bb} which depend on the worldsheet coordinates.}. Additional simplifications 
come from the fact 
that bosonic propagators in the AdS light-cone gauge  are only diagonal, which limitates  the 
number of Feynman graphs to be considered~\footnote{In the first two-loop calculation of \cite{Roiban:2007ju} 
the conformal gauge was used, 
in which propagators are non-diagonal, implying the evaluation of a larger number of two-loop 
diagrams.}.  In general,  the actual calculation inherits from its $AdS_5\times S^5$ a similar mechanism of 
cancellation of divergences, and even the significative difference given by the 
presence of massless fermions in the spectrum turns out not to play a role (apart from the cancellation of UV divergences) in our final result, as they behave like effectively decoupled.
The relevant interaction vertices are the same and  no 
genuinely new contributions, in terms of scalar integrals, appears. This results 
in a different weight factor  in 
front of the same structures ($\log 2$ at one loop and the Catalan constant $K$ at two loops) 
appearing in the $AdS_5\times S^5$ case, where the weight is in terms of the ratio of the $AdS_4$ and 
$\cp^3$ radii, as well as the number of bosonic transverse AdS directions and 
massive fermions.

%The complicated structure of the GS action makes the verification of cancellation 
%of UV divergences (power-like, ln2 Î and ln Î ones) non- trivial at the 2-loop order.
% 

An important further ingredient in the \adscp calculation is the correction
to the effective string tension~\cite{Bergman:2009zh} which must be considered for the
first time at  this 
order in sigma-model perturbation theory.
The  original 
``dictionary'' 
proposal~\cite{Aharony:2008ug} for the effective string tension  in terms of the effective 't Hooft 
coupling $\lambda$ of ABJM reads 
\be\label{shift}
T= 
\frac{R^2}{2\pi\alpha'}
=2\sqrt{2 \lambda}\,,
\qquad\qquad
\lambda=\frac{N}{k}\,,
\ee
where $R$ is the $\cp^3$ radius. As pointed out in \cite{Bergman:2009zh}, the geometry 
(and flux, in the ABJ \cite{Aharony:2008gk} theory)  of the background induces higher order corrections 
to  the radius of curvature in the Type IIA description, which in the planar 
limit %~\footnote{The anomalous 
%radius shift  in the Type 
%IIA description of \adscp\cite{Bergman:2009zh}, due to 
%effects of curvature corrections and discrete torsion, is in general given in 
%terms of units of  torsion (3-form) flux quanta and orbifold parameter $k>1$.} 
of interest here appear in the form of a shift in the square root
\begin{equation}\label{eq:tension}
T = 2\sqrt{2\left(\lambda-\frac{1}{24}\right)}\,.
\end{equation}
We emphasize that the string perturbative expansion is  an expansion in inverse string tension 
whose coefficients are obviously not affected by the correction~\eqref{eq:tension}. 
The radius shift is a (corrected) AdS$_4$/CFT$_3$ dictionary proposal, an assumed, new input which plays a role 
when expressing the result in terms of the 't Hooft coupling.

All this leads to the main result of this work, which is the evaluation of  the first two strong coupling 
corrections to the ABJM cusp anomalous dimension
\begin{equation}\label{cusp_result}
f_{\abjm}(\lambda) = \sqrt{2\lambda } - \frac{5 \log 2}{2 \pi } - \left(\frac{K}{4\pi^2} + 
\frac{1}{24}\right)\frac{1}{\sqrt{2\lambda}} + {\cal O}(\sqrt{\lambda})^{-2}
\,.
\end{equation}
The formula can be rewritten in a more compact way defining the shifted coupling
\begin{equation}
\tilde\lambda \equiv \lambda - \frac{1}{24}\,,
\end{equation}
from which
\begin{equation}
f_{\abjm}\left(\tilde\lambda\right) = \sqrt{2\tilde\lambda } - \frac{5 \log 2}{2 \pi } - 
\frac{K}{4\pi^2\,\sqrt{2\tilde\lambda}} + {\cal O}(\sqrt{\tilde\lambda})^{-2}\,.
\end{equation}
This form of the result makes evident the striking similarity with the  $AdS_5\times S^5$ 
result
\be
f_{\YM} (\lambda_\YM) = \frac{\sqrt{\lambda_\YM }}{\pi} - \frac{3 \log 2}{ \pi } 
- \frac{K}{\pi\,\sqrt{\lambda_\YM}} 
+ {\cal O}(\sqrt{\lambda_\YM})^{-2}\,,
\ee 
where  the change in the transcendentality pattern is due to the corresponding difference
in the effective string tensions. %(at variance with the $AdS_5\times S^5$ case, where 
%$T=\sqrt{\lambda}/(2\pi)$, here the tension \eqref{eq:tension} has transcendentality zero).

From \eqref{cusp_result} and via \eqref{cusp_pred} we get then the strong-coupling two-loop correction for the interpolating 
function $h(\lambda)$, that we report here together with the weak coupling results
~\cite{Gaiotto:2008cg,Grignani:2008is,Nishioka:2008gz,Minahan:2009aq,Minahan:2009wg,Leoni:2010tb}
\begin{equation}\label{eq:prediction}
\begin{array}{lll}
h^2(\lambda)& = \displaystyle 
\lambda^2 - \frac{2\,\pi^3}{3}\,\lambda^4+ {\cal O}\left(\lambda^6\right) & \quad \lambda \ll 1\quad , \\
h(\lambda)& = \displaystyle\sqrt{\frac{\lambda}{2}} - \frac{\log2}{2\pi} - \frac{1}{48\sqrt{2\lambda}} 
+ {\cal O}(\sqrt{\lambda})^{-2} & \quad \lambda \gg 1 \quad,
\end{array}
\end{equation}
where we emphasize the  a priori non-obvious fact the two-loop coefficient at strong coupling is 
only due to the  anomalous radius shift. 

A conjecture for the exact expression of $h(\lambda)$
has  been recently made~\cite{Gromov:2014eha}, in a spirit quite close to the one followed
in~\cite{Correa:2012at,Correa:2012hh} on the comparison 
between two exact
computations of the same observable (see footnote 1).
The authors of ~\cite{Gromov:2014eha} elaborated  on the similarity between two all-order 
calculations in ABJM theory: one 
- the ``slope function''~\cite{Basso:2011rs}~-
derived via integrability as exact solution of a quantum spectral curve ~\cite{Cavaglia:2014exa} 
and one - a 1/6 BPS Wilson loop~\cite{Kapustin:2009kz,Marino:2009jd,Drukker:2010nc}
 - obtained with supersymmetric localization. 
%and based on previous experience in $\mathcal{N}=4$ with the so-called 
%Bremsstrahlung function~\cite{Drukker:2012de,Correa:2012hh,Gromov:2012eu}. 
As the first of the two exact results is expressed in terms of the 
effective coupling $h(\lambda)$,  an  ``extrapolation'' for the latter has been derived in an 
exact,  implicit, form~\footnote{As noticed in~\cite{Gromov:2014eha}, a more solid derivation of 
$h(\lambda)$ would require comparison between the localization results 
of~\cite{Marino:2009jd,Drukker:2010nc} and  the ABJM 
Bremsstrahlung function~\cite{Griguolo:2012iq,Lewkowycz:2013laa, Bianchi:2014laa, Correa:2014aga}, similarly to the case of  
the $h(\lambda_\YM)$ of $\mathcal{N}=4$ SYM, see footnote 1.}.
It is 
\begin{equation}\label{eq:proposal}
\lambda = \frac{\sinh{2\pi h(\lambda)}}{2\pi}\, _3F_2 \left(\frac12,\frac12,\frac12 ; 1,\frac32 ; 
-\sinh^2{2\pi h(\lambda)} \right) \,,
\end{equation}
with weak and strong coupling expansions 
\begin{align} 
h(\lambda)&= \lambda-\frac{\pi^2}{3}\,\lambda^3+\frac{5\pi^4}{12}\,\lambda^5-
\frac{893\pi^6}{1260}\,\lambda^7+\mathcal{O}(\lambda^9)  & \lambda &\ll 1 \, , 
\\\label{eq:strong}
h(\lambda) &=  \sqrt{\frac12\left( \lambda - \frac{1}{24}\right)} - \frac{\log 2}{2 \pi}  
+ {\cal O}\left( e^{-2\pi\sqrt{2\lambda}} \right)
& \lambda &\gg 1 \, .
\end{align}
We see that  \eqref{eq:strong} above, expanded for large $\lambda$,  agrees with 
\eqref{eq:prediction}.  
 
In general, the mutual consistency of several ingredients - our direct perturbative string calculation, 
the corrected dictionary of~\cite{Bergman:2009zh},  the prediction \eqref{cusp_pred}-\eqref{cusp_pred_2} 
from the Bethe Ansatz~\cite{Gromov:2008qe} 
and  the conjecture of~\cite{Gromov:2014eha} for the interpolating function 
$h(\lambda)$  - provides highly non-trivial evidence in support of the proposal  
\eqref{eq:proposal} for the interpolating function $h(\lambda)$ of ABJM theory, 
and furnishes an indirect check of the quantum integrability of the \adscp 
superstring theory in this $\kappa$-symmetry light-cone gauge.

%we will find match the strong-coupling Bethe ansatz [23] predictions [24, 25]; 
%this provides evidence for the quantum integrability of the superstring formulated in the

The paper proceeds as follows. In Section \ref{sec:ads_lc} we introduce the $AdS$ light-cone gauge-fixed action
which in Section \ref{sec:fluctuation} we write in terms of fluctuations over the null cusp classical solution. % of~\cite{Giombi:2009gd}. 
In Section  \ref{sec:oneloop} we compute %expand around the null cusp background and reproduce the same small fluctuation spectrum
%found in the semiclassical analysis of folded spinning strings  \cite{McLoughlin:2008he,Abbott:2010yb,LopezArcos:2012gb}, thus leading 
the 
%same 
one-loop correction to the cusp anomaly.
In Section \ref{sec:twoloops} we  extend the computation of the string partition function to 
one more order, verifying the cancellation of UV divergences and obtaining the strong coupling 
two-loop correction to the ABJM cusp anomaly. %, similarly to the $AdS_5\times S^5$ case,  
%in terms of the Catalan constant.  
 In Appendix \ref{app:a} we present for completeness a different  parametrization of the 
 $\kappa$-symmetry gauge-fixed action of~\cite{Uvarov:2009nk} which can be  transparently 
 compared with its $AdS_5\times S^5$ counterpart. Appendices \ref{app:lagr_exp} and \ref{app:intred}
contain, respectively, details on the expanded Lagrangian and
 explicit reductions for the relevant integrals which we use in Section 
 \ref{sec:twoloops}.
 
%%%%%%%%%%%%%%%%%%%%%%%%%%%%%%%%%
\section{AdS light-cone gauge in $AdS_4\times\mathbb{CP}^3$}
\label{sec:ads_lc}
%%%%%%%%%%%%%%%%%%%%%%%%%%%%%%%%%

Our starting point is the $AdS_4\times \mathbb{CP}^3$ Lagrangian in the $\kappa$-symmetry 
light-cone gauge proposed in \cite{Uvarov:2009hf,Uvarov:2009nk}. 
This is obtained by double dimensional reduction from the eleven-dimensional membrane action~\cite{deWit:1998yu} based on the supercoset 
    $OSp(8|4)/\left(SO(7)\times SO(1, 3)\right)$, 
 and choosing a $\kappa$-symmetry light-cone gauge for which both light-like directions lie in 
 $AdS_4$. 
 In the spirit of
~\cite{Metsaev:2000yf, Metsaev:2000yu} (and of earlier studies 
 of brane models on the $AdS\times S$ backgrounds)  the construction  of~\cite{Uvarov:2009hf, Uvarov:2009nk} 
 formulates the bulk string theory in a way 
which is naturally related to the boundary CFT theory. In particular, the 32-dimensional spinors
whose components 
are the coordinates  associated to the odd generators of $OSp(8|4)$
are divided  
in $\theta$ and $\eta$ fermions corresponding, respectively,  to super-Poincar\'e generators   
and superconformal generators. The AdS $\kappa$-symmetry light-cone gauge consists in setting to zero 
that half of the fermions which correspond to fermionic generators having  negative charge w.r.t. the $SO(1,1)$ 
 generator $M^{+-}$ from the Lorentz group acting on the Minkowski boundary of 
 $AdS_4$% Namely, $\Gamma^+\Theta=(\Gamma^0+\Gamma^3)\Theta=0$, 
%where $\Theta$ comprises the 
%32 
% fermionic coordinates  associated to the odd generators of $OSp(8|4)$. The gauge-fixing sets to zero 
% half of the fermionic coordinates, 
%where ``half'' is defined with respect to the $SO(1,1)$ rotations in light-cone 
%directions~\cite{Metsaev:2000yf}
~\footnote{Another $\kappa$-symmetry gauge condition based on a similar ``superconformal'' basis
has been considered in~\cite{Zarembo:2009au}.}. 
As our analysis below explicitly shows, it has the advantage of encompassing a quantum analysis of string 
configurations classically moving in the $AdS_4$ sector of 
 $AdS_4\times \mathbb{CP}^3$~\footnote{An alternative  $\kappa$-symmetry  gauge fixing of the complete 
\adscp superspace~\cite{Gomis:2008jt} which is suitable for studying regions of the theory 
that are not reachable by the supercoset sigma model of~\cite{Arutyunov:2008if, Stefanski:2008ik} 
(see Introduction) has been considered 
in~\cite{Grassi:2009yj}.}.

The $AdS_4\times \mathbb{CP}^3$ background metric is
\begin{equation}\label{eq:metric}
ds^2_{10} = R^2 \left( \frac14\, ds^2_{AdS_4} + ds^2_{\mathbb{CP}^3} \right)\,,
\end{equation}
where $R$ is the $\cp^3$ radius. For $AdS_4$  the Poincar\'e patch is used  and 
 the parametrization of  $\cp^3$ is at this stage arbitrary 
\begin{eqnarray}
ds^2_{AdS_4} &=& \frac{dw^2 + dx^+ dx^- + dx^1dx^1}{w^2} \qquad\qquad x^{\pm} \equiv x^2 \pm 
x^0\,,\\
ds^2_{\cp^3}&=&g_{MN}\,dz^M dz^N\,\qquad\qquad M=1,...,6~.
\end{eqnarray}
Above, $x^\pm$ are the light-cone coordinates, $x^m=(x^0,x^1,x^2)$ parametrize the 
three-dimensional  boundary of $AdS_4$ and $w\equiv e^{2\varphi}$ is the radial coordinate.
%The $\mathbb{CP}^3$ part is parametrized by six coordinates $z^M$, $M=1,\dots 6$.\\
The $\kappa$-symmetry light-cone gauge-fixed Lagrangian of~\cite{Uvarov:2009hf, Uvarov:2009nk} 
can be written as follows~\footnote{Inspired by \cite{Metsaev:2000yf} we modify the action proposed in \cite{Uvarov:2009hf,Uvarov:2009nk} with a convenient rescaling of the fermions 
\begin{equation}
\theta_{a}  \rightarrow \sqrt{2}\, \theta_{a} \qquad
\theta_{4} \rightarrow \sqrt{2}\, e^{-\varphi}\theta_{4} \qquad
\eta_{a} \rightarrow \sqrt{2}\, e^{-2\varphi}\eta_{a} \qquad
\eta_{4} \rightarrow \sqrt{2}\, e^{-\varphi}\eta_{4}
\end{equation}
and similar ones for the complex conjugates. With respect to~\cite{Uvarov:2009hf, Uvarov:2009nk}, 
we also partially change notation.}
\begin{align}\label{eq:lagrangian}
S &= -\frac{T}{2} \int\, d\tau\, d\sigma\, L \qquad\qquad \\ %T = \frac{R^2}{2\pi\alpha'}\\ 
L &= \gamma^{ij}\Big[\frac{e^{-4\varphi}}{4} \left(\partial_ix^+\partial_jx^-+
\partial_ix^1\partial_jx^1 \right) + \partial_i\varphi
\partial_j\varphi + g_{MN}\partial_i z^M \partial_j z^N\nonumber\\&+
e^{-4\varphi}\left(\partial_ix^+\varpi_j+\partial_ix^+\partial_jz^M h_M+e^{-4\vf} B\partial_ix^+\partial_jx^+\right)\Big]\nonumber\\
&-2\, \varepsilon^{ij}e^{-4\varphi}\left({\omega}_i\partial_jx^++e^{-2\varphi}C\partial_ix^1\partial_jx^+
+\partial_ix^+\partial_jz^M\ell_M\right)\nonumber \quad ,
\end{align}
where the string tension $T$ has been defined in \eqref{eq:tension} and   the following quantities
\begin{align}\label{eq:pieces}
\varpi_i&=i\left(\partial_i\theta_a\bar{\theta}^a-\theta_a\partial_i\bar{\theta}^a
+\partial_i\theta_4\bar{\theta}^4-\theta_4\partial_i\bar{\theta}^4
+\partial_i\eta_a\bar{\eta}^a-\eta_a\partial_i\bar{\eta}^a
+\partial_i\eta_4\bar{\eta}^4-\eta_4\partial_i\bar{\eta}^4\right) \, ,\\ 
\omega_i&=\hat{\eta}_a\hat{\partial_i}\bar{\theta}^a
+\hat{\partial_i}\theta_a\hat{\bar{\eta}}^a+\frac12\left(\partial_i\theta_4\bar{\eta}^4-\partial_i\eta_4
\bar{\theta}^4+\eta_4\partial_i\bar{\theta}^4-\theta_4
\partial_i\bar{\eta}^4\right) \, , \\ \label{eq:B}
B&=8\,  \left[(\hat{\eta}_a\hat{\bar{\eta}}^a)^2+\varepsilon_{abc}\hat{\bar{\eta}}^a\hat{\bar{\eta}}^b
\hat{\bar{\eta}}^c\bar{\eta}^4+\varepsilon^{abc}
\hat{\eta}_a\hat{\eta}_b\hat{\eta}_c\eta_4+2
\eta_4\bar{\eta}^4 \left(\hat{\eta}_a\hat{\bar{\eta}}^a-\theta_4\bar{\theta}^4\right)\right] \, , \\ 
C&=2\, \hat{\eta}_a\hat{\bar{\eta}}^a+
\theta_4\bar{\theta}^4+\eta_4\bar{\eta}^4 \, ,\\ 
h_M&=2\, \left[\Omega^a_M\varepsilon_{abc}\hat{\bar{\eta}}^b\hat{\bar
{\eta}}^c-\Omega_{aM}\varepsilon^{abc}\hat{\eta}_b
\hat{\eta}_c + 2\left(\Omega_{aM}\hat{\bar{\eta}}^a
\bar{\eta}^4-\Omega^a_M\hat{\eta}_a\eta_4\right) + 2
\left(\theta_4\bar{\theta}^4+\eta_4\bar{\eta}^4\right)
\tilde{\Omega}_{a\ M}^{\ a}\right]\, , \\ \label{eq:ellM}
\ell_M&=2\, i\, \left[\Omega_{aM}\hat
{\bar{\eta}}^a\bar{\theta}^4+\Omega^a_M\hat{\eta}_a\theta_4
+ \left(\theta_4\bar{\eta}^4-\eta_4\bar{\theta}^4\right)
\tilde{\Omega}_{a\ M}^{\ a}\right] 
\end{align}
include fermions up to the fourth power. As in the $AdS_5\times S^5$ case~\cite{Metsaev:2000yf, Metsaev:2000yu}, the action is quadratic  in the $\theta$-fermions  and quartic in the $\eta$-fermions.

Above, the fermionic coordinates $\eta_a$ and $\theta_a$ (and their conjugates) 
transform in the fundamental (antifundamental) representation of $SU(3)$ ($a=1,2,3$), and 
correspond to the unbroken 24 supersymmetries of the \adscp background.
The remaining fermions $\eta_4$, $\theta_4$ and their conjugates originate 
from the eight broken supersymmetries. 
The manifest symmetry of the action 
is thus only the $SU(3)$ subgroup of the $SU(4)$  global symmetry of 
$\cp^3$. 
This feature, as we will see, will be inherited by the quantum fluctuations around the light-like 
cusp (see also discussion in Appendix \ref{app:a}). 
The $\Omega^a_M$ and $\Omega_{aM}$ appearing in the Lagrangian are the complex vielbein 
of $\mathbb{CP}^3$, $ds^2_{\mathbb{CP}^3} = \Omega^a_M \Omega_{aN}\, dz^M\, 
dz^N$, namely components of the Cartan one-forms of $SU(4)/U(3)$,  
$\Omega^a = \Omega^a_M\, dz^M$ and $\Omega_a = \Omega_{aM}\, dz^M$. 
In the construction of~\cite{Uvarov:2009hf}, $\tilde\Omega_a^{\phantom{a}a} $  
 is associated to a one-form corresponding to the fiber direction of $S^7$. Its expression is given explicitly below in terms of the $\cp^3$ coordinates. 
The  $\Omega^a_M$ and $\tilde\Omega_a^{\phantom{a}a} $  appear in~\cite{Uvarov:2009hf} in a  
``dressed'' $OSp(6|4)/(U(3)\times SO(1, 3))$ supercoset element where the 
dressing incorporates the information on the broken supersymmetries and $U(1)$ fiber direction.
In \eqref{eq:pieces}, hatted quantities are related to unhatted ones via a 
rotation by  matrices $T$ (similar matrices were conveniently introduced in~\cite{Metsaev:2000yu}) 
which depend on the $\cp^3$ coordinates and act
 as follows on e.g. a $\eta_a$ fermion
\begin{equation}\label{eq:T}
\hat\eta_a = T_a^{\phantom{a}b}\, \eta_b + T_{ab}\,\bar\eta^b\, , \qquad\qquad \hat{\bar\eta}^a 
= T^a_{\phantom{a}b}\, \bar\eta^b + T^{ab}\,\eta_b\,.
\end{equation}
In Appendix \ref{app:a} we rewrite the Lagrangian \eqref{eq:lagrangian} in a form that is more similar
to the $AdS_5\times S^5$ of \cite{Metsaev:2000yf}, and comment more on the Cartan forms $\Omega$ and
$T$-matrices. 

The parametrization for $\mathbb{CP}^3$ chosen in~\cite{Uvarov:2008yi} consists of complex variables $z^a$ and $\bar z_a$, 
transforming in the $\mathbf{3}$ and $\mathbf{\bar 3}$ of $SU(3)$ respectively.
Then the metric reads
\begin{equation}
ds^2_{\mathbb{CP}^3} = g_{ab}\, dz^a\, dz^b + g^{ab}\, d\bar z_a\, d\bar z_b + 2\, 
g_a^{\phantom{a}b}\, dz^a\,d\bar z_b\,,
\end{equation}
where 
\begin{align}
& g_{ab} = \frac{1}{4|z|^4} \left( |z|^2 - \sin^2{|z|} + \sin^4{|z|} \right)\bar z_a\,\bar z_b\, , \qquad g^{ab} = \frac{1}{4|z|^4} \left( |z|^2 - \sin^2{|z|} + \sin^4{|z|} \right) z^a\, z^b\, , & \nonumber\\
& g_a^{\phantom{a}b} = \frac{\sin^2{|z|}}{2|z|^2}\, \delta_a^b + \frac{1}{4|z|^4} \left( |z|^2 - \sin^2{|z|} - \sin^4{|z|} \right)\bar z_a\, z^b \qquad \text{and}\quad |z|^2 \equiv z^a\, \bar z_a\,. &
\end{align}
For the one-forms  appearing in the Lagrangian explicit expressions then follow, which 
can be derived from their definition 
\begin{align}
\Omega^a = \Omega^a_{\phantom{a},b}\, dz^b + \Omega^{a,b}\, d\bar z_b\, , \qquad \Omega_a = \Omega_{a,b}\, d z^b + \Omega_a^{\phantom{a},b}\, d\bar z_b \, ,
\qquad \tilde\Omega_a^{\phantom{a}a} 
= \tilde \Omega_{a\phantom{a},b}^{\phantom{a}a}\, dz^b + \tilde \Omega_a^{\phantom{a}a,b}\, d\bar 
z_b\, \, ,
\end{align}
using \eqref{eq:forms}. For example, 
\begin{equation}\label{omtildecc}
\tilde \Omega_a^{\phantom{a}a} = i\, \frac{\sin^2|z|}{|z|^2} \left( dz^a\, \bar z_a - z^a\, d\bar z_a 
\right)\,.
\end{equation}
In this parametrization, the T-matrices introduced in \eqref{eq:T} can be grouped in a unitary matrix 
${T_{\hat{a}}}^{\hat b}$ which reads explicitly~\cite{Uvarov:2008yi} 
\begin{equation}\label{matrixT}
{T_{\hat{a}}}^{\hat b}=\left(
 \begin{array}{cc}
  T_a^{\phantom{a}b} & T_{ab} \\
  T^{ab} & T^a_{\phantom{a}b}
 \end{array}\right) %= \left(
% \begin{array}{cc}
%  0 & i\,\varepsilon_{acb}\, z^c \\
 % -i\,\varepsilon^{acb}\, \bar z_c & 0
 %\end{array}\right)= 
= \left(
 \begin{array}{cc}
  \delta_a^b\, \cos{|z|} + \bar z_a\, z^b\, \frac{1-\cos{|z|}}{|z|^2} & i\,\varepsilon_{acb}\, z^c\, \frac{\sin{|z|}}{|z|} \\
  -i\,\varepsilon^{acb}\, \bar z_c\, \frac{\sin{|z|}}{|z|} & \delta^a_b\, \cos{|z|} + z^a\, \bar z_b\, \frac{1-\cos{|z|}}{|z|^2}
 \end{array}\right)\,.
\end{equation} 
The action \eqref{eq:lagrangian} has gauge-fixed local fermionic symmetry.
To fix bosonic local symmetry and further proceed with our analysis it is convenient to use, as 
discussed in~\cite{Metsaev:2000yf} and used in~\cite{Giombi:2009gd, Giombi:2010fa, 
Giombi:2010zi}, a ``modified'' conformal gauge
 \begin{equation}\label{eq:nonconf}
\gamma^{ij} = {\rm diag}\left(-e^{4\varphi}, e^{-4\varphi}\right)\,,
\end{equation}
in combination with the standard   light-cone gauge 
\be\label{eq:lcdir}
x^+=p^+\,\tau\,,\qquad p^+={\rm const}\,.
\ee
%\begin{align*}
%L & =\Big\{\frac{1}{4}\left[-\left(\dot{x}^{1}\right)^{2}+e^{-8\varphi}\left(x^{1'}\right)^{2}\right]-e^{4\varphi}\left(\dot{\varphi}^{2}+g_{MN}\dot{z}^{M}\dot{z}^{N}\right)+e^{-4\varphi}\left(\varphi^{'2}+g_{MN}z^{M'}z^{N'}\right)\\
% & -p^{+}\left[\varpi_{\tau}+\dot{z}^{M}h_{M}+p^{+}e^{-4\varphi}B\right]\Big\} +2p^{+}e^{-4\varphi}\left(\tilde{\omega}_{\sigma}+e^{-2\varphi}C\partial_{\sigma}x^{1}-z^{M'}\ell_{M}\right)
%\end{align*}
In what follows we will give directly the expression of the Euclidean version of the action  \eqref{eq:lagrangian}
 in this gauge (choosing $p^+=1$)  and on the null cusp background~\cite{Kruczenski:2002fb, Giombi:2009gd}.

%%%%%%%%%%%%%%%%%%%%%%%%%%%%%%%%%
\section{The null cusp fluctuation action}
\label{sec:fluctuation}
%%%%%%%%%%%%%%%%%%%%%%%%%%%%%%%%%

In this section we consider the Wick-rotated, Euclidean formulation of the Lagrangian \eqref{eq:lagrangian} 
in the bosonic light-cone gauge \eqref{eq:nonconf}-\eqref{eq:lcdir} and compute its fluctuations about the null cusp background.
The equations of motion derived from the (Euclidean) AdS light-cone gauge Lagrangian \eqref{eq:lagrangian} 
admit a classical solution for which the on-shell action is the area of the minimal surface 
ending on a null cusp on the $AdS_4$ boundary.
This configuration is just the $AdS_4$ embedding of the classical string solution found in the $AdS_5$ 
background~\cite{Kruczenski:2002fb, Giombi:2009gd}, and reads
\begin{eqnarray}\label{eq:background}
& w \equiv e^{2\varphi} = \displaystyle\sqrt{\frac{\tau}{\sigma}} \qquad\qquad
x^{1} = 0 &\nonumber \\
& x^{+} =\tau \qquad\qquad
x^{-} =-\displaystyle\frac{1}{2\sigma}  \qquad\qquad
z^{M} = 0~. &
\end{eqnarray} 
The requirement that the open string Euclidean world-sheet described by these coordinates ends on a 
cusp at the boundary of $AdS_4$ at $w=0$ is manifestly enforced by the relation $x^+\, x^- = -\frac12 w^2$.
%This is obtained from  the $AdS_5 \times S^5$ case and obtained from it by just removing one $AdS_5$ coordinate (say $x^2$ in \cite{Giombi:2009gd}). We refer to the abundant literature on the cusp in $AdS_5\times S^5$ for further explanations \cite{Kruczenski:2002fb,Roiban:2007jf,Kruczenski:2007cy,Roiban:2007dq,Giombi:2009gd}.
In the AdS/CFT dictionary of \cite{Maldacena:1998im,Rey:1998ik}, the Wilson loop evaluated on a light-like cusp contour is then given by the superstring partition function
\begin{equation}\label{eq:partition}
\left\langle W_{cusp} \right\rangle = Z_{string} \equiv \int {\cal D}[x,w,z,\theta,\eta]\, e^{-S_E}\,.
\end{equation}
In order to compute it perturbatively, we first construct the Euclidean action $S_E$ for 
fluctuations about the background \eqref{eq:background}. 
Following~\cite{Giombi:2009gd}, we will use a suitable parametrization of fluctuations which, combined with 
a further redefinition of the worldsheet coordinates  
$t = \log \tau$ and $s=\log\sigma$, is such  that  
the coefficients of the fluctuation action become constant, namely 
$(\tau,\sigma)$-independent. It reads~\footnote{The factor 2 in the fluctuation of the field $x^1$ is
 introduced to normalize the kinetic term of $\tilde x^1$.}
\begin{eqnarray}\label{eq:fluctuations}
& x^1 = 2\, \displaystyle\sqrt{\frac{\tau}{\sigma}} \tilde x^1 \qquad\qquad w = \displaystyle\sqrt{\frac{\tau}{\sigma}}\, \tilde w \qquad\qquad \tilde w = e^{2\tilde\varphi}  & \nonumber\\
& z^a = \tilde z^a \qquad\qquad \bar z^a = \tilde{\bar z}^a \qquad\qquad a=1,2,3 & \nonumber\\
& \eta = \displaystyle\frac{1}{\sqrt{\sigma}}\, \tilde \eta \qquad\qquad \theta = \displaystyle\frac{1}{\sqrt{\sigma}}\, \tilde \theta\,. & 
\end{eqnarray}
After the Wick rotation $\tau\to -i\,\tau, p^+\to i p^+$ and having set $p^+=1$, 
we end up    with the following action for  fluctuations over the 
null-cusp background \eqref{eq:background}
\begin{equation}\label{eq:Lagrangian_exp}
S_E = \frac{T}{2}\, \int dt\, ds\, {\cal L}\qquad,\qquad
{\cal L} = {\cal L}_B + {\cal L}_F^{(2)} + {\cal L}_F^{(4)}\,,
\end{equation}
where
\begin{align}
 {\cal L}_B& = \left( \partial_t \tilde x^1 + \frac12\, \tilde x^1 \right)^2 + \frac{1}{\tilde w^4} \left( \partial_s \tilde x^1 - \frac12 \tilde x^1 \right)^2
+ \tilde w^2\, \left(\partial_t \varphi \right)^2 + \frac{1}{\tilde w^2}\, \left(\partial_s \varphi \right)^2 + \frac{1}{16} \left( \tilde w^2 + \frac{1}{\tilde w^2} \right) + 
\nonumber\\& 
+ \tilde w^2 \, \tilde g_{MN}\, \partial_t \tilde z^M\, \partial_t \tilde z^N + \frac{1}{\tilde w^{2}}\, \tilde g_{MN}\, \partial_s \tilde z^M\, \partial_s \tilde z^N
\\
{\cal L}_F^{(2)} &= i\Big[  \partial_t \tilde{\theta}_a \tilde{\bar\theta}^a - \tilde{\theta}_a\partial_t\tilde{\bar\theta}^a
+ \partial_t\tilde{\theta_4}\tilde{\bar\theta}^4 - \tilde{\theta_4}\partial_t
\tilde{\bar\theta}^4 + \partial_t\tilde{\eta_a}\tilde{\bar\eta}^a - \tilde{\eta_a}\partial_t\tilde{\bar\eta}^a 
+ \partial_t\tilde{\eta_4}\tilde{\bar\eta}^4 - \tilde{\eta_4}\partial_t
\tilde{\bar\eta}^4 \Big] + \nonumber\\&
+\frac{2i}{w^2}\Big[\hat{\eta}_a \left(\hat{\partial_s}\bar{\theta}^a - \frac12\, \hat{\bar \theta}^a \right)
+ \left(\hat{\partial_s}\theta_a - \frac12\, \hat \theta_a \right) \hat{\bar{\eta}}^a + \frac12 \left(\partial_s\theta_4\bar{\eta}^4 - \partial_s\eta_4
\bar{\theta}^4 + \eta_4\partial_s\bar{\theta}^4 - \theta_4
\partial_s\bar{\eta}^4\right)\Big]\nonumber\\&
+ \partial_t \tilde z^M\, \tilde h_M + \frac{4\,i}{\tilde w^3}\, \tilde C\, \left( \partial_s \tilde x^1 
- \frac12 \tilde x^1 \right) - \frac{2i}{\tilde w^2}\, \partial_s \tilde z^M\, \tilde\ell_M  
\\\label{eq:Lagrangian_exp_fin}
{\cal L}_F^{(4)} &= \frac{1}{\tilde w^4}\, \tilde B \,.
\end{align}
In the expressions above,  with $\tilde B$, $\tilde C$, $\tilde h_M$ and $\tilde \ell_M$ we indicate the
quantities $ B$, $  C$, $  h_M$ and $ \ell_M$ in \eqref{eq:pieces} where a tilde 
over each field appears (namely, the weighting factors for the fluctuations  
in \eqref{eq:fluctuations} have already been made explicit in the derivatives of products). 
%The additional factors $i$ in the last three terms of ${\cal L}_F^{(2)}$ 
%come from Wick-rotating the original action.

Since the Lagrangian has now constant coefficients and is thus translationally invariant, 
the (infinite) world-sheet volume factor $V$ factorizes. The scaling function is then defined 
via the string partition function as~\cite{Giombi:2009gd}
\be\label{Wpert}
W=-\ln Z=\frac{1}{2}f(\lambda) \,V=W_0+W_1+W_2+... \,,
\qquad\qquad V=\frac{1}{4}V_2\equiv\frac{1}{4} \int dt\, ds
\ee
where $W_0\equiv S_E$ coincides with the value of the action on the background, 
$W_1, W_2, ...$ are one-, two- and higher loop corrections, and for the ratio $V/V_2$ 
we use the same  convention as in~\cite{Giombi:2009gd}~\footnote{This is related to coordinate 
transformation and field redefinitions occurring  between the GKP~\cite{Gubser:2002tv} string, whose energy is given in terms of $f(\lambda)$, and the null cusp solution in the Poincar\'e patch here used, see discussion in~\cite{Giombi:2010zi}. }.
%is due to the coordinate transformations used above%~\footnote{
%The rescaling $t = \log \tau$ and $s=\log\sigma$ of the world-sheet coordinates, which 
%puts the induced worldsheet metric in the conformal gauge 
%$ds^2_{\rm ind}=\frac{1}{4}(dt^2+ds^2)$. The 2-dimensional volume is then $V=\frac{1}{4}\int dt 
%ds$.} 
From \eqref{Wpert} we explicitly define $f(\lambda)$ in terms of the 
effective action $W$
\begin{equation}\label{eq:cuspprescription}
f(\lambda) = \frac{8}{V_2}\, W\, .
\end{equation}
We are now ready to compute the effective action perturbatively 
in inverse powers of the effective string tension $g \equiv \frac{T}{2}$. From 
this we will extract the corresponding strong coupling perturbative expansion for the scaling function
\begin{equation}\label{eq:expansion}
f(g) = g\, \left[ 1 + \frac{a_1}{g} + \frac{a_2}{g^2} + \dots 
\right]\,,\qquad\qquad g=\frac{T}{2}~.
\end{equation} 
where we have factorized the classical result from $W_0 = S_E$~\cite{Aharony:2008ug} 
and the effective string tension $T$ is defined in \eqref{eq:tension}.

%%%%%%%%%%%%%%%%%%%%%%%%%%%%%%%%%
\section{Cusp anomaly at one loop}\label{oneloop}
\label{sec:oneloop}
%%%%%%%%%%%%%%%%%%%%%%%%%%%%%%%%%
We start considering one-loop quantum corrections to the free energy \eqref{eq:partition}, 
which are derived expanding the fluctuation Lagrangian \eqref{eq:Lagrangian_exp} to second order in the fields.\\
For the bosonic part we obtain
\begin{equation}
\mathcal{L}_{B}^{(2)} = \left(\partial_t \tilde x^1\right)^2 + \left(\partial_s \tilde x^1\right)^2 + \frac{1}{2}\left(\tilde{x}^1\right)^2 + \left(\partial_t\tilde\varphi\right)^2 + \left(\partial_s\tilde\varphi\right)^2 + \tilde\varphi^2 + \left|\partial_t \tilde z^a\right|^2 + \left|\partial_s \tilde z^a\right|^2\,.
\end{equation}
The bosonic degrees of freedom consist of six real massless scalars (associated to the $\mathbb{CP}^3$ coordinates), one real scalar $\tilde x^1$ with mass $m^2 = \frac12$ and one real scalar $\tilde\varphi$ with mass $m^2 = 1$.
This is a simple truncation  (one less transverse degree of freedom in the AdS space)  of the bosonic spectrum
found in the $AdS_5\times S^5$~\cite{Giombi:2009gd}.
%Up to the number of scalar fields of each type, this is the same bosonic spectrum as in $AdS^5 \times S^5$.\\
For the fermions one gets an off-diagonal kinetic matrix
\begin{equation}
{\cal L}^{(2)}_F = i\, \Theta\, K_{F}\, \Theta^{T} \quad \text{where} \quad \Theta \equiv \left(\tilde{\theta}_{a},\tilde{\theta}_{4},\tilde{\bar{\theta}}^{a},\tilde{\bar{\theta}}^{4},\tilde{\eta}_{a},\tilde{\eta}_{4},\tilde{\bar{\eta}}^{a},\tilde{\bar{\eta}}^{4}\right)\,,
\end{equation}
which reads
\begin{equation}\label{eq:KF}
K_F=
\begin{pmatrix}
0 & 0 & -\partial_t & 0 & 0 & 0 & -\partial_s-\frac{1}{2} & 0 \\
0 & 0 & 0 & -\partial_t  & 0 & 0 & 0 & -\partial_s \\ 
-\partial_t  & 0 & 0 & 0 & \partial_s+\frac{1}{2} & 0 & 0 & 0 \\
0 & -\partial_t  & 0 & 0 & 0 & \partial_s & 0 & 0 \\
0 & 0 & \partial_s -\frac{1}{2} & 0 & 0 & 0 & -\partial_t  & 0 \\
0 & 0 & 0 & \partial_s & 0 & 0 & 0 & -\partial_t  \\
-\partial_s+\frac{1}{2} & 0 & 0 & 0 & -\partial_t  & 0 & 0 & 0 \\
0 & -\partial_s & 0 & 0 & 0 & -\partial_t  & 0 & 0\\
\end{pmatrix}\,.
\end{equation}
Fermions contribute to the partition function with the determinant 
($\partial_{\mu} = i\, p_\mu\,,\,\mu=0,1 $)
\begin{equation}
\text{det}\ K_F = \left(p^2\right)^2\left(p^2+\frac{1}{4}\right)^6\,,
\end{equation}
from which we read that the fermionic spectrum is composed of six massive degrees of freedom with mass $m^2=1/4$ and two massless ones. The latter  are of $\eta_4$ and $\theta_4$ type, namely  those fermionic directions corresponding to the broken supersymmetries.
The presence of massless fermions marks a difference with respect to the ${\cal N}=4$ SYM case,
already noticed in this theory when studying fluctuations over classical  string 
solutions only lying in $AdS_4$~\cite{McLoughlin:2008he,Abbott:2010yb,LopezArcos:2012gb, Forini:2012bb}
(see comments in section \ref{comparison}).

The one-loop effective action is computed as
\begin{equation}
W_1 = -\log Z_1
\end{equation}
where $Z_1$ is the ratio of fermionic over bosonic determinants. Therefore
\begin{equation}
W_1 = \frac{1}{2}\, V_2\, \int{\frac{d^2p}{(2\pi)^2}\left[\log \left(p^2+1\right)+\log \left(p^2+\frac{1}{2}\right) + 4\log\left(p^2\right) - 6\log\left(p^2+\frac{1}{4}\right)\right]} = -\frac{5\log 2}{16\pi}\, V_2\,.
\end{equation}
The one-loop correction to the scaling function reads, according to \eqref{eq:cuspprescription},
\begin{equation}\label{eq:a1}
a_1 = -\frac{5\log 2}{2\pi}
\end{equation}
and agrees with previous independent results \cite{McLoughlin:2008he,Abbott:2010yb,LopezArcos:2012gb}.

%%%%%%%%%%%%%%%%%%%%%%%%%%%%%%%%
\section{Cusp anomaly at two loops}\label{twoloops}
\label{sec:twoloops}
%%%%%%%%%%%%%%%%%%%%%%%%%%%%%%%%
In this section we provide the details on the computation of the two-loop coefficient of the scaling function. The calculation follows the lines of \cite{Giombi:2009gd}, with some important differences which we point out in section \ref{comparison}. 
In particular the aim is to compute the connected vacuum diagrams of the fluctuation Lagrangian around the null cusp background. Denoting by $W$ the free energy of the theory, $W=-\log Z$, the two-loop contribution is given by
\begin{equation}
 W_2=\braket{S_{int}}-\frac12\braket{S_{int}^2}_c\,,
\end{equation}
where $S_{int}$ is the interacting part of the action at cubic and quartic order (see appendix \ref{expansion}). 
The subscript $c$ indicates that only connected diagrams need to be included. In the following we use $S_{int}=T\int\, dt\, ds\,  \mathcal{L}_{int}$ and we give the expressions of the vertices as they appear in $\mathcal{L}_{int}$. Throughout this section we drop tildes from fluctuation fields in order not to clutter formulae. Also,  we neglect the string tension $T$ and the volume $V_2$ in the intermediate steps and reinstate them at the end of the calculation.

\subsection{Bosonic sector}
Let us first consider the purely bosonic sector. 
As pointed out in section \ref{oneloop}, the spectrum of the theory contains one real boson of squared mass 1, one real boson of squared mass $\frac12$ and three complex massless bosons. The interaction among these excitations involves cubic and quartic vertices which give rise to the diagrams in figure \ref{diagrams}.
\begin{figure}[h]\label{diagrams}
\begin{center}
\begin{tikzpicture}[scale=1, vertex/.style={circle,fill=black,thick,inner sep=2pt}]
\draw(-5,0) circle (1 cm);
\node (c) at (-6,0) [vertex] {};
\node (c) at (-4,0) [vertex] {};
 \draw[-] (-6,0)--(-4,0);
\draw(-1,0) circle (1 cm);
\draw(1,0) circle (1 cm);
 \node (c) at (0,0) [vertex] {};
 \draw[-] (5.5,0)--(6.5,0);
 \draw(7.5,0) circle (1 cm);
\draw(4.5,0) circle (1 cm);
\node (c) at (5.5,0) [vertex] {};
\node (c) at (6.5,0) [vertex] {};
 \end{tikzpicture}
\end{center}
\caption{Sunset, double bubble and double tadpole are the diagrams appearing in the two-loop contribution to the partition function.}
\end{figure}
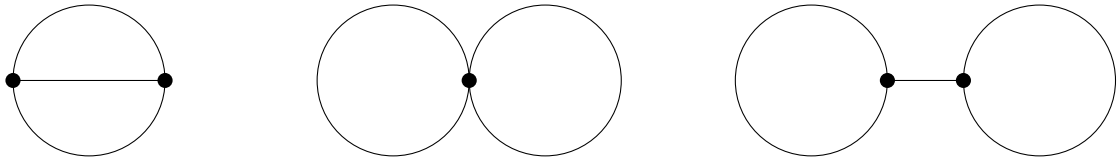\\
We observe that the AdS light-cone gauge Lagrangian contains only diagonal bosonic propagators, which introduces considerable simplifications in the perturbative computation. The explicit expressions of the propagators are
\begin{equation}
G_{\varphi\varphi}(p) = \frac{1}{p^2+1}\qquad\qquad G_{z_a \bar z^b}(p) = \frac{2\, \delta_a^b}{p^2}\qquad\qquad G_{x^1x^1}(p) = \frac{1}{p^2+\frac12}\,.
\end{equation}
The cubic interactions involving only bosonic fields are of three different kinds
\begin{equation}\label{bos3ver}
 V_{\varphi x^1 x^1}=-4\,\varphi\, \left[(\pa_s-{\textstyle\frac12})\, x^1\right]^2 \qquad V_{\varphi^3}=2\varphi\left[(\pa_t \varphi)^2-(\pa_s\varphi)^2\right] \qquad V_{\varphi |z|^2}=2\varphi\left[|\pa_t z|^2-|\pa_s z|^2\right]\,.
\end{equation}
When combining vertices and propagators in the sunset diagrams they originate various non-covariant integrals with components of the loop momenta in the numerators. Standard reduction techniques allow to rewrite every integral as a linear combination of the two following scalar ones (explicit reductions for the relevant integrals are spelled out in appendix \ref{app:intred})
\begin{align}
I\left(m^2\right) & \equiv \int \frac{d^2p}{\left(2\pi\right)^2}\, \frac{1}{p^2+m^2} \\
I\left(m_1^2,m_2^2,m_3^2\right) & \equiv \int \frac{d^2p\, d^2q\,d^2r}{\left(2\pi\right)^4}\,  \frac{\delta^{(2)}(p+q+r)}{(p^2+m_1^2)(q^2+m_2^2)(r^2+m_3^2)} \quad. 
\end{align}
The latter integral is finite, provided none of the masses vanishes, and is otherwise IR divergent.
The former is clearly UV logarithmically divergent, and also develops IR singularities in the massless case. 
In our computation we expect all UV divergences to cancel and therefore no divergent integral to appear in the final result.
Nonetheless, performing reduction of potentially divergent tensor integrals to scalar ones still implies the choice of a regularization scheme. In our case we use the one adopted in \cite{Giombi:2009gd,Roiban:2007jf,Roiban:2007dq}.
This prescription consists of performing all manipulations in the numerators in $d=2$, which has the advantage of simpler tensor integral reductions.
In this process we set to zero power UV divergent massless tadpoles, as in dimensional regularization 
\begin{equation}
\int \frac{d^2p}{(2\pi)^2}\, \left( p^2\right)^n = 0\,, \qquad\qquad n\geq 0.
\end{equation}
All remaining logarithmically divergent integrals happen to cancel out in the computation and there is no need to pick up an explicit regularization scheme to compute them.\\
As an explicit example, we consider   the contribution to the sunset coming from the first vertex in \eqref{bos3ver}
\begin{equation}\label{sunsetxx}
-\frac12\braket{V_{\varphi x^1 x^1}^2}= -\int \frac{d^2p\, d^2q\,d^2r}{(2\pi)^4}\,\frac{(1+4 q_1^2)\, (1+4 r_1^2)\, \delta^{(2)}(p+q+r)}{(p^2+1) (q^2+\frac12) (r^2+\frac12)} =\frac12\, I\left(1,{\textstyle\frac12},{\textstyle\frac12}\right)\,.
\end{equation}
The reason why the coefficient of the integral in the second term of \eqref{sunsetxx} is exactly $(-1)$ is the topic of section \ref{comparison}. We note that the integral $I\left(1,{\textstyle\frac12},{\textstyle\frac12}\right)$ already appeared in \cite{Giombi:2009gd} and is a particular case of the general class
\begin{equation}\label{intcat}
 I\left(2\,m^2,m^2,m^2\right)=\frac{K}{8\, \p^2\, m^2}\,,
\end{equation}
where $K$ is the Catalan constant
\begin{equation}
K \equiv \sum_{n=0}^{\infty} \frac{(-1)^n}{(2n+1)^2} \,.
\end{equation}
The contribution of the sunset diagram involving the second vertex in \eqref{bos3ver} is proportional to $I(1)^2$, whereas the contribution of the third vertex vanishes
\begin{equation}
-\frac12\braket{V_{\varphi^3}^2}=2\, I(1)^2  \qquad\qquad -\frac12\braket{V_{\varphi |z|^2}^2}=0
\end{equation}
The final contribution of the bosonic sunset diagrams is
\begin{equation}\label{bossun}
 W_{2,\textup{bos. sunset}}=\frac12\, I\left(1,{\textstyle\frac12},{\textstyle\frac12}\right)+2\,I(1)^2\,.
\end{equation}
The first two vertices in \eqref{bos3ver} can also be contracted to generate non-1PI graphs, 
namely double tadpoles. However the resulting diagrams turn out to vanish individually.

Next we consider bosonic double bubble diagrams. The relevant quartic vertices are 
\begin{align}
& V_{\varphi^2 x^1 x^1}=16\,\varphi^2\, \left[(\pa_s-{\textstyle\frac12})\, x^1\right]^2 \label{phixx} \\
& V_{\varphi^4}=4\, \varphi^2\left[(\pa_t \varphi)^2+(\pa_s\varphi)^2+\frac16\varphi^2\right] \label{phi4}\\
& V_{\varphi^2 |z|^2}=4\, \varphi^2 \left[|\pa_t z|^2+|\pa_s z|^2\right]\\
& V_{z^4}=\frac16 \left[ \left(\bar z_a \pa_t z^a\right)^2 + \left(\bar z_a \pa_s z^a\right)^2 + \left(z^a \pa_t \bar z_a\right)^2 + \left(z^a \pa_s \bar z_a\right)^2
\right.\nonumber\\& \hspace{1cm} \left.
- |z|^2\left(|\pa_t z|^2+|\pa_s z|^2\right) - \left|\bar z_a \pa_t z^a\right|^2 - \left|\bar z_a \pa_s z^a\right|^2\right]\,.
\end{align}
Despite the lengthy expressions of the vertices, the only non-vanishing contribution comes from $V_{\varphi^4}$ and gives
\begin{equation}
 W_{2,\textup{bos. bubble}}=-2\, I(1)^2~
\end{equation}
and cancels the divergent part of \eqref{bossun}.
As a result, the bosonic sector turns out to be free of divergences without the need of fermonic contributions,
which was already observed in the $AdS_5\times S^5$ case~\cite{Giombi:2009gd}.

%%%%%%%%%%%%%%%%
\subsection{Fermionic contributions}\label{fermcontr}
%%%%%%%%%%%%%%%%
We compute the diagrams arising from interactions involving fermions.
The fermionic propagators can be read from the inverse of the kinetic matrix $K_F$ \eqref{eq:KF}
\begin{align}
 G_{\eta_4\bar\eta^4}(p) &= G_{\theta_4\bar\theta^4}(p) = \frac{p_0}{p^2} &
 G_{\eta_4\bar\theta^4}(p) &= G_{\theta_4\bar\eta^4}(-p) = -\frac{p_1}{p^2}\nonumber \\
 G_{\eta_a\bar\eta^b}(p) &= G_{\theta_a\bar\theta^b} (p)= \frac{p_0}{p^2+\frac14}\delta_a^b &   G_{\eta_a\bar\theta^b}(p) &=G_{\theta_a\bar\eta^b}(-p)= -\frac{p_1+\frac{i}{2}}{p^2+\frac14}\delta_a^b 
\end{align}
The main difference between the spectrum of $AdS_5\times S^5$ and the one introduced in section \ref{oneloop} resides in the fermionic part. Although both theories have eight fermionic degrees of freedom, in $AdS_4\times\mathbb{CP}^3$ they are split into six massive and two massless excitations, which interact non-trivially among themselves.\\
We start by considering diagrams involving at least one massless fermion.
The relevant cubic vertices are (we denote by $\psi$ the fermions $\eta$ and $\theta$ collectively)
\begin{align}\label{fer3verm0}
 V_{z\eta_a\eta_4}&=\ -2\, \pa_t z^a \eta_a \eta_4 + h.c.&
 V_{z\eta_a\theta_4}&=2\, \pa_s z^a \eta_a \theta_4-h.c. \nonumber\\
 V_{\varphi \eta_4 \bar\theta^4}&= -2\,i\, \varphi\, (\bar\theta^4 \pa_s \eta_4-\pa_s\bar\theta^4 \eta_4)-h.c.&
 V_{x^1 \bar\psi^4 \psi_4}&= -2\, i\, (\bar\eta^4 \eta_4+\bar\theta^4 \theta_4) (\pa_s-{\textstyle\frac12}) x^1\,.
\end{align}
The quartic interactions are either not suitable for constructing a double tadpole diagram or they produce vanishing integrals. These include vector massless tadpoles, which vanish by parity, and tensor massless tadpoles, which have power UV divergences and are set to zero. For completeness we list them in appendix \ref{expansion}. \\
Focussing on the Feynman graphs which can be constructed from cubic interaction 
we also note that the only double tadpole diagrams that can be produced using \eqref{fer3verm0} involve tensor
massless tadpole integrals and therefore vanish.
In the sector with massless fermions we are therefore left with the sunset diagrams, which, thanks to the diagonal structure of the bosonic propagators, turn out to be only five
\begin{equation}
W_{2,\psi_4}=-\frac12 \braket{V_{z\eta_a\eta_4}V_{z \eta_a\eta_4}+V_{z\eta_a\theta_4}V_{z\eta_a\theta_4}+2\,V_{z\eta_a\eta_4}V_{z\eta_a\theta_4}+ V_{\varphi \eta_4 \bar\theta^4} V_{\varphi \eta_4 \bar\theta^4}+ V_{x^1 \bar\psi^4 \psi_4} V_{x^1 \bar\psi^4 \psi_4}}\,.
\end{equation}
The explicit computation of the individual contributions shows that they are all vanishing. As an example we consider 
\begin{equation}
-\frac12 \braket{V_{\varphi \eta_4 \bar\theta^4} V_{\varphi \eta_4 \bar\theta^4}}=4 \int \frac{d^2p\, d^2q\,d^2r}{(2\p)^4} \frac{(p_1-q_1)^2(p_0q_0-p_1q_1)\, \delta^{(2)}(p+q+r)}{p^2 q^2 (r^2+1)} = 0
\end{equation}
and similar cancellations happen for the other diagrams. Therefore we conclude that  $W_{2,\psi_4}=0$ and that massless fermions are effectively decoupled at two loops.\\
We then move to consider massive fermions, starting from their cubic coupling to bosons
\begin{align}
 V_{z\eta\eta}&= -\e^{abc} \pa_t \bar z_a \eta_b \eta_c + h.c.&
 V_{z\eta\theta}&= -2\, \e^{abc} \bar z_a \eta_b (\pa_s-{\textstyle\frac12})\theta_c-h.c. \nonumber\\
 V_{\varphi\eta \theta}&= -4\, i\, \varphi\, \eta_a (\pa_s-{\textstyle\frac12})\bar\theta^a -h.c. & 
 V_{x^1 \eta \eta}&= -4\, i\, \bar \eta^a \eta_a (\pa_s-{\textstyle\frac12}) x^1 \,.
\end{align}
Precisely as in the massless case, this generates five possible sunset diagrams. None of them is vanishing. We present the details of a particularly relevant example, i.e. the one involving the vertex  $V_{x^1 \eta \eta}$. This gives
\begin{align}\label{fermsun1}
 -\frac12 \braket{V_{x^1 \eta \eta} V_{x^1 \eta \eta}} =24 \int \frac{d^2p\, d^2q\,d^2r}{(2 \pi)^4}\, \frac{(p_1^2+{\textstyle\frac14})\,q_0\,r_0\, \delta^{(2)}(p+q+r)}{(p^2+{\textstyle\frac12})(q^2+{\textstyle\frac14})(r^2+{\textstyle\frac14})} = -\frac{3}{8}\,  I\left({\textstyle\frac12},{\textstyle\frac14},{\textstyle\frac14}\right) + \frac34\, I\left({\textstyle\frac14}\right)^2\,.
\end{align}
We note the appearance of another integral in the class \eqref{intcat}. The coefficient in front of this integral depends on the degrees of freedom of the theory and is thoroughly discussed in section \eqref{comparison}. The partial results of the remaining sunset diagrams are
\begin{align}\label{fermsun2}
& -\frac12 \braket{(V_{z\eta\eta}+V_{z\eta\theta})(V_{z\eta\eta}+V_{z\eta\theta})}= 3\, I\left({\textstyle\frac14}\right)^2 - 6\, I\left({\textstyle\frac14}\right)\, I(0)\nonumber\\
& -\frac12 \braket{ V_{\varphi\eta \theta}  V_{\varphi\eta \theta} }_{\textup{1PI}}=6\, I\left({\textstyle\frac14}\right)\, I(1)+\frac34\, I\left({\textstyle\frac14}\right)^2\,.
\end{align}
The latter vertices can be contracted also in a non-1PI manner
\begin{equation}
-\frac12 \braket{ V_{\varphi\eta \theta}  V_{\varphi\eta \theta} }_{\textup{non-1PI}}=-\frac12\, G_{\varphi\varphi} (0)\times 2^6 \times 3^2\times \int \frac{d^2 p}{(2\p)^2}\, \frac{p_1^2+\frac14}{p^2+\frac14}=-\frac92\, I\left({\textstyle\frac14}\right)^2
\end{equation}
where the factor in front of the integrals comes from the expression of the vertex and from counting the degrees of freedoms that can run in the loops. 
As in \cite{Giombi:2009gd}, the divergent contribution proportional to $I\left({\textstyle\frac14}\right)^2$ cancels exactly those coming from \eqref{fermsun1} and \eqref{fermsun2}.\\ 
The total cubic fermionic part reads
\begin{equation}
 W_{2,\textup{ferm. cubic}}=-\frac{3}{8}\,  I\left({\textstyle\frac12},{\textstyle\frac14},{\textstyle\frac14}\right)+6\, I\left({\textstyle\frac14}\right)\,I(1) - 6\, I\left({\textstyle\frac14}\right)\, I(0)\,.
\end{equation}
Finally we consider the fermionic double bubble diagrams. These involve the fermionic quartic vertices. However, most of the vertices appearing in the Lagrangian cannot contribute to the partition function either because the bosonic propagators are diagonal or because they would produce vanishing integrals. We present the whole list of quartic vertices in appendix \ref{expansion} and we spell out here only the relevant ones for our computation
\begin{align}
 V_{\varphi^2 \eta \theta}&= 8\, i\, \varphi^2\, \eta_a (\pa_s-{\textstyle\frac12})\bar\theta^a -h.c. &
 V_{zz\eta\theta}&=-2\, i\, \left[|z|^2 \eta_a (\pa_s-{\textstyle\frac12})\bar\theta^a - \bar z_b z^a\eta_a (\pa_s-{\textstyle\frac12})\bar\theta^b\right] -h.c.\,.
\end{align} 
Although we can build a diagram with $ V_{\eta^4}$, fermion propagators carry one component of the loop momentum in the numerator and produce vector tadpole integrals, which vanish by parity. We conclude that the contribution from fermionic double bubble graphs is
\begin{equation}
W_{2,\textup{ferm. bubbles}}=-6\, I\left({\textstyle\frac14}\right)\,I(1) + 6\, I\left({\textstyle\frac14}\right)\, I(0)\,.
\end{equation}

Summing all the partial results and reinstating the dependence on the string tension and the volume, we obtain
\begin{equation}\label{finalres}
 W_2=\frac{V_2}{T} \left[\frac12\, I\left(1,{\textstyle\frac12},{\textstyle\frac12}\right)-\frac38 I\left({\textstyle\frac12},{\textstyle\frac14},{\textstyle\frac14}\right)\right] =-\frac14 \frac{V_2}{T} I\left(1,{\textstyle\frac12},{\textstyle\frac12}\right)=-\frac{K}{16\, \pi^2}\frac{V_2}{T}
\end{equation}
where $T$ is defined in \eqref{eq:tension}.
Finally we can plug this expression into equation \eqref{eq:cuspprescription} and read out the second order of the strong coupling expansion \eqref{eq:expansion} of the ABJM cusp anomalous dimension
\begin{equation}\label{eq:a2}
a_2 = -\frac{K}{4\pi^2}\,.
\end{equation}
%%%%%%%%%%%%%%%%%%%%%%%%%%%%%%%%
\subsection{The cusp anomalous dimension}
%%%%%%%%%%%%%%%%%%%%%%%%%%%%%%%%
We summarize the results of our superstring computation, presenting the strong coupling expansion of the ABJM cusp anomalous dimension up to two-loop order.
Reinstating the definition of the string tension \eqref{eq:tension} in terms of the ABJM 't Hooft coupling and plugging \eqref{eq:a1} and \eqref{eq:a2} into \eqref{eq:expansion}, we find
\begin{equation}
f_{ABJM}(\lambda) = \sqrt{2\lambda } - \frac{5 \log 2}{2 \pi } - \left(\frac{K}{4\pi^2} + \frac{1}{24}\right)\frac{1}{\sqrt{2\lambda}} + {\cal O}\left(\lambda^{-1}\right) \,,
\end{equation}
which is the main result of the paper.
From the string dual point of view it looks convenient to define the shifted coupling
\begin{equation}
\tilde\lambda \equiv \lambda - \frac{1}{24}\,,
\end{equation}
in terms of which we can rewrite the scaling function more compactly as
\begin{equation}
f_{ABJM}\left(\tilde\lambda\right) = \sqrt{2\tilde\lambda } - \frac{5 \log 2}{2 \pi } - \frac{K}{4\pi^2\,\sqrt{2\tilde\lambda}} + {\cal O}\left(\tilde\lambda^{-1}\right)\,.
\end{equation}

%%%%%%%%%%%%%%%%%%%%%%%%%%%%%%%%
\subsection{Comparison with $AdS_5\times S^5$}\label{comparison}
%%%%%%%%%%%%%%%%%%%%%%%%%%%%%%%%
In this section we point out similarities and differences between the calculation we performed and 
its $AdS_5\times S^5$ analogue \cite{Giombi:2009gd}. 
The starting points, i.e. the Lagrangians in $AdS$ light-cone gauge, look rather different. Yet the final results of the two-loop computations are strikingly similar.
More precisely, when written in terms of the string tension, the two expressions have exactly the same structure up to the numerical coefficients in front of the integrals.
Indeed the $AdS_5$ computation gives 
\begin{equation}\label{resAdS5}
 W^{(AdS_5)}_2=\frac{V_2}{T} \left[\frac14\, I\left(1,{\textstyle\frac12},{\textstyle\frac12}\right)-\frac14\, I\left({\textstyle\frac12},{\textstyle\frac14},{\textstyle\frac14}\right)\right]\,,
\end{equation}
which looks very similar in structure to \eqref{finalres}.
Furthermore, using \eqref{intcat}, both combinations sum up to 
\begin{equation}
W_2=-\frac{V_2}{T}\,\frac14\, I\left(1,{\textstyle\frac12},{\textstyle\frac12}\right)
\end{equation}
and only the different relation between the string tension and the 't Hooft couplings distinguishes the final results.
It is easy to trace the origin of the integrals and their coefficients back in the vertices of the Lagrangian and to understand their meaning. 
In particular in both computations only the sunset diagrams involving the interactions $V_{\varphi xx}$ and $V_{x\psi\psi}$ (with massive fermions) seem to effectively contribute. All other terms are also important, but just serve to cancel divergences.
Hence we can now focus on the relevant interactions and point out the differences between the $AdS_5$ and the $AdS_4$ cases. \\
We start from the bosonic sectors.
The two theories differ for the number of scalar degrees of freedom with given masses.
Focussing on massive fluctuations, after gauge fixing we have one scalar with $m^2=1$ associated to the radial coordinate of $AdS_{d+1}$ and $(d-2)$ real scalars with $m^2=\frac12$.
In the metric we chose for the $AdS_4\times \mathbb{CP}^3$ background, the size of the $AdS_4$ part is rescaled by a factor of $r^2=4$. We have compensated this, parametrizing the radial coordinate as $w=e^{r\varphi}$ and introducing a factor $r$ in the fluctuation of $x^1$, so as to have the same normalization for their kinetic terms as in $AdS_5\times S^5$. This causes some factors $r$ to appear in interaction vertices in our Lagrangian. Apart from this, the relevant interaction vertices are exactly the same.
Then, the number of $x$ fields $(d-2)$ and this factor $r$ determine the coefficient of the integral $I\left(1,{\textstyle\frac12},{\textstyle\frac12}\right)$ appearing in equations \eqref{finalres} and \eqref{resAdS5}.\\
Turning to fermions, the first striking difference between the $AdS_5$ and $AdS_4$ cases is the presence of massless ones.
As pointed out at the beginning of section \ref{fermcontr} their contribution is effectively vanishing at two loops (though they do contribute at first order).
Focussing on massive fermions, the relevant cubic interactions giving rise to $I\left({\textstyle\frac12},{\textstyle\frac14},{\textstyle\frac14}\right)$ look again similar in the $AdS_4$ and $AdS_5$ cases. The difference is given once more by the ratio of the radii $r$ (through the normalization of $\varphi$ and $x$ coordinates) and the number $n_f$ of massive fermions in the spectrum ($n_f=8$ for $AdS_5\times S^5$ and $n_f=6$ for $AdS_4\times \mathbb{CP}^3$).\\ 
The final results \eqref{finalres} and \eqref{resAdS5} can be re-expressed in the general form
\begin{align}\nonumber
 W^{(AdS_{d+1})}_2&=\frac{V_2}{T} \frac{(d-2)r^2}{8} \left[I\left(1,{\textstyle\frac12},{\textstyle\frac12}
 \right)-\frac{n_f}{8}\, I\left({\textstyle\frac12},{\textstyle\frac14},{\textstyle\frac14}\right)\right]\\\label{finalresgen}
 &=\frac{V_2}{T} \frac{(d-2)r^2}{8} \left(1-\frac{n_f}{4}\right)\, I
 \left(1,{\textstyle\frac12},{\textstyle\frac12}\right)\,,\qquad \qquad d=3,4\,,
\end{align}
where the cases at hand are $d=4$, $n_f=8$, $r=1$ for ${\cal N}=4$ SYM and $d=3$, $n_f=6$, $r=2$ for ABJM.
 
\section{Concluding remarks}
%%%%%%%%%%%%%%%%%%%%%%%%%%%%%%%%%
In this work we have computed the cusp anomalous dimension of ABJM theory up to second order in its 
strong coupling  expansion. 
This result has been determined considering  the $AdS_4 \times \cp^3$ $\kappa$-symmetry gauge-fixed
action of~\cite{Uvarov:2009hf, Uvarov:2009nk} and studying its fluctuations about the null cusp background 
\eqref{eq:background}, which is a classical solution thereof. 
As in the $AdS_5\times S^5$ counterpart of this calculation~\cite{Giombi:2009gd}, the $AdS$ light-cone gauge 
approach~\cite{Metsaev:2000yf}  makes the explicit evaluation rather manageable, 
allowing us to push the expansion of the string partition function up to second order. 

While at one loop we confirm a known result 
\cite{McLoughlin:2008he,Abbott:2010yb,LopezArcos:2012gb}, 
at two-loops we provide a new important piece of data, see \eqref{cusp_result}, which we combine with 
a proposal based on the 
Bethe Ansatz of $AdS_4$/CFT$_3$~\cite{Gromov:2008qe} to give our two-loop 
correction to the so-called interpolating function $h(\lambda)$ of ABJM theory, 
equation \eqref{eq:prediction}.  
Importantly, the recent conjecture of~\cite{Gromov:2014eha} for an all-order 
expression  of $h(\lambda)$, implicitly given in terms of a non-trivial hypergeometric function,
 agrees with our result, which is a relevant perturbative test  of validity 
 for the conjecture.  
 In particular, at this level of perturbation theory we must implement in our calculation a ``correction'' 
to the string tension in terms of the 't Hooft  coupling, which was pointed out 
in~\cite{Bergman:2009zh} to be due to higher order corrections (in curvature) 
to the background. We show that  the strong coupling two-loop 
correction for $h(\lambda)$ is  only due to the anomalous  shift of the curvature radius
in the  Type IIA  description~\cite{Bergman:2009zh}. This supports the 
observation in \cite{Gromov:2014eha}  on the origin of the shift appearing in its proposal 
\eqref{eq:strong}, which knows nothing about the gravity side but coincides in fact with the correction of
~\cite{Bergman:2009zh}.

In perspective, we observe that the light-cone gauge approach could be pushed to a much stronger 
check of \eqref{eq:proposal}, by testing its finite coupling regime.
Following \cite{McKeown:2013vpa}, one could discretize the light-cone Lagrangian 
\eqref{eq:Lagrangian_exp}, put it on a lattice and perform numerical simulations to determine 
the ABJM scaling function in terms of the coupling constant, for any value thereof.
By comparison with the same results for ${\cal N}=4$ SYM one could then provide 
numerical values of $h(\lambda)$ at some finite values of $\lambda$, which could 
then be contrasted with \eqref{eq:proposal}.

The manifest cancellation of  UV divergences that we find here provides a 
direct demonstration of the quantum consistency of the \adscp action 
of~\cite{Uvarov:2009hf,Uvarov:2009nk}, and shows that it can be readily used for non-trivial 
strong coupling computations in the $AdS_4/$CFT$_3$ framework (following for 
example~\cite{Giombi:2010fa, Giombi:2010zi}).  In particular, the 
consistency of the result with predictions coming from integrability, the 
 conjecture~\cite{Gromov:2014eha} and the ``corrected'' dictionary of~\cite{Bergman:2009zh} can be taken as evidence, albeit indirect, of quantum integrability for 
the Type IIA \adscp superstring in this gauge.

%%%%%%%%%%%%%%%%%%%%%%%%%%%%%%%%%
%%%%%%%%%%%%%%%%%%%%%%%%%%%%%%%%%
\section*{Acknowledgments}
%%%%%%%%%%%%%%%%%%%%%%%%%%%%%%%%%
It is a pleasure to thank Alessandra Cagnazzo, Andrea Cavagli\`a, Ben Hoare, 
Valentina G.M.Puletti,  Nikolay Gromov, Radu Roiban, Roberto Tateo and Arkady Tseytlin 
 for discussions, and in particular Ben Hoare for useful comments on the draft. 
The work of LB, AB, VF and EV is funded by DFG via the Emmy Noether Program
``Gauge Fields from Strings''. 
%%%%%%%%%%%%%%%%%%%%%%%%%%%%%%%%%

\setcounter{equation}{0}

%%%%%%%%%%%%%%%%%%%%%%%%%%%%%%%%%
\appendix
%%%%%%%%%%%%%%%%%%%%%%%%%%%%%%%%%

%%%%%%%%%%%%%%%%%%%%%%%%%%%%%%%%%
\section{Lagrangian in the Wess-Zumino type parametrization}
%%%%%%%%%%%%%%%%%%%%%%%%%%%%%%%%%
\label{app:a}

%The light-cone gauge action can be found in two related forms. 
%One of them corresponds to the Wess-Zumino type gauge in 10-d superspace 
%while another is based on the Killing gauge (see [3, 10]). 
%These âgaugesâ (better to be called âparametrizationsâ) do not reduce the number 
%of fermionic degrees of freedom but only specialize a choice of fermionic coordinates. 
%The action given in this Section corresponds to the WZ parametrization, 
%while the action in the Killing parametrization will be discussed in Section 6.

%The light-cone gauge action can be found in two related forms. 
%One of them corresponds to the Wess-Zumino type gauge in 10-d superspace 
%while another is based on the Killing gauge (see [3, 10]). 
%These ÒgaugesÓ (better to be called ÒparametrizationsÓ) do not reduce the number 
%of fermionic degrees of freedom but only specialize a choice of fermionic coordinates. 
%The action given in this Section corresponds to the WZ parametrization, 
%while the action in the Killing parametrization will be discussed in Section 6.

In this appendix we rewrite the Lagrangian \eqref{eq:lagrangian} in a form that resembles the 
Wess-Zumino type parametrization introduced in \cite{Metsaev:2000yf}, and compare it to the $AdS_5\times S^5$ case. In \cite{Metsaev:2000yf} the authors found two possible ways to eliminate the fermion rotation \eqref{eq:T}, either by a change of parametrization for $S^5$ or by the introduction of a covariant derivative for the terms quadratic in fermions. Here we explore only the second option and we leave the first one for future development.
We first introduce a collective index for upper and lower indices so that  
\begin{equation} 
 \eta_{\hat{a}}=\left(\begin{array}{c}
                      \eta_a \\
                      \bar\eta^a 
                      \end{array}\right)\,.\, 
\end{equation}
In this notation the action of the matrix $T$ on the fermions \eqref{eq:T} can be rewritten as 
\begin{equation}
\hat{\eta}_{\hat a}= {T_{\hat{a}}}^{\hat b}\eta_{\hat b}
\end{equation}
where the matrix ${T_{\hat{a}}}^{\hat b}$ is given in \eqref{matrixT}. 
We also  introduce the shorthand notation
\begin{equation}
 \partial_i\eta_a\bar{\eta}^a-\eta_a\partial_i\bar{\eta}^a=-\eta^{\hat{a}}\partial_i\eta_{\hat{a}}~,
\end{equation}
where $\eta^{\hat{a}}=(\bar{\eta}^a,\eta_a)$. In \cite{Metsaev:2000yf} a recipe for going from the 
Killing parametrization to a Wess-Zumino type gauge was given, which consists of rotating back the fermions. 
This generates additional terms coming from derivatives that can be reabsorbed into a covariant derivative. 
In particular, we apply the transformation
\begin{equation}
 \eta_{\hat a}\to \big(T^{-1}\big)_{\hat a}^{\hat b} \,\eta_{\hat b}\,.
\end{equation}

In contrast with the $AdS_5\times S^5$ case the matrix $T$ is not block diagonal, therefore one has $\eta^{\hat{a}}\partial_i\eta_{\hat{a}}=\hat{\eta}^{\hat{a}}\hat\partial_i {\eta}_{\hat{a}}$, where 
it is crucial to use  hatted indices. This transformation removes all the hats from fermions, at the price of introducing the covariant derivative 
\begin{equation}
 D=d-\Omega\,,
\end{equation}
where $\Omega\equiv{\Omega_{\hat a}}^{\hat b}=d {{T}_{\hat a}}^{\hat c}\, {(T^{-1})_{\hat c}}^{\hat b}$ and $d\Omega-\Omega\wedge\Omega=0$. More explicitly\footnote{
The matrix $\Omega$ was already introduced in \cite{Uvarov:2008yi} 
however there it was defined as 
${\Omega_{\hat a}}^{\hat{b}}=i {T_{\hat a}}^{\hat c}d  {{T^{-1}}_{\hat c}}^{\hat b}
=-i d{T_{\hat a}}^{\hat c} {{T^{-1}}_{\hat c}}^{\hat b}$, differing from ours by a factor of $i$. 
To make contact with the expressions of \cite{Uvarov:2008yi} 
we add such a factor in formula \eqref{omega}.}, the (matrix) Cartan form  
entering the definition of the (dimensionally reduced) supercoset element reads
\begin{equation}\label{omega}
 {\Omega_{\hat a}}^{\hat{b}}=i\, \left(\begin{array}{cc}
                              \Omega_{a}^{\phantom{a}b} - \d_a^b \Omega_c^{\phantom{c}c}& \e_{acb}\Omega^c\\
                              -\e^{acb}\Omega_{c} & -\Omega^a_{\phantom{a}b}+\d_b^a \Omega_c^{\phantom{c}c}
                             \end{array}\right)\,,
\end{equation}
with components given by  
\begin{align}\label{eq:forms}
\Omega_{a}^{\phantom{a}b}&=i\frac{(1-\cos{|z|})}{|z|^2}(\bar z_adz^b-d\bar z_az^b)-i\bar z_az^b\frac{(1-\cos{|z|})^2}{2|z|^4}(dz^c\bar z_c-z^cd\bar z_c),\\
\Omega_{a}&=d\bar z_a\frac{\sin{|z|}}{|z|}+\bar z_a\frac{\sin{|z|}(1-\cos{|z|})}{2|z|^3}(dz^c\bar z_c-z^cd\bar z_c)+\bar z_a\left(\frac{1}{|z|}-\frac{\sin{|z|}}{|z|^2}\right)d|z|,\\
\Omega^{a}&=dz^a\frac{\sin{|z|}}{|z|}+z^a\frac{\sin{|z|}(1-\cos{|z|})}{2|z|^3}(z^cd\bar
z_c-dz^c\bar
z_c)+z^a\left(\frac{1}{|z|}-\frac{\sin{|z|}}{|z|^2}\right)d|z|.
\end{align}
Above, $\Omega_c^{\phantom{a}c}$ is the trace of \eqref{eq:forms} and is related 
to $\tilde\Omega_c^{\phantom{a}c}$ defined in \eqref{omtildecc} via 
$\tilde\Omega_c^{\phantom{a}c}=2\,\Omega_c^{\phantom{a}c}$.

We can also decompose the matrix $\Omega$ in order to separate the contributions from the vielbein and from the spin connection\footnote{A similar procedure was applied in \cite{Metsaev:2000yf} where in that case the decomposition is expressed in terms of the $SO(5)$ $\gamma$-matrices.} 
\begin{equation}\label{decomposition}
 {\Omega_{\hat a}}^{\hat{b}}=\Omega^{\hat c} {(E_{\hat c})_{\hat a}}^{\hat b}+ \Omega^c_{\phantom{c}d} {(J_c^d)_{\hat a}}^{\hat b}
\end{equation}
with\footnote{Let us stress that the meaning of the first term of equation \eqref{decomposition} in matrix form is the following 
\begin{equation}
\Omega^{\hat c}{(E_{\hat c})_{\hat a}}^{\hat{b}}=\left(\begin{array}{cc}
                             \Omega^c{(E_{c})_a}^b+\Omega_c{(E^{c})_a}^b&  \Omega^c (E_{c})_{ab}+\Omega_c (E^{c})_{ab}\\
                               \Omega^c (E_{c})^{ab}+\Omega_c(E^{c})^{ab} &  \Omega^c{(E_{c})^a}_b+\Omega_c{(E^{c})^a}_b
                             \end{array}\right)
\end{equation}
and the explicit expression of ${(E_{\hat c})_{\hat a}}^{\hat{b}}$ shows that the only non-vanishing elements are $(E_{c})_{ab}$ and $(E^{c})^{ab}$.}
\begin{align}
 {(E_{\hat c})_{\hat a}}^{\hat{b}}&=i\, \left(\begin{array}{cc}
                             0& \e_{acb}\\
                              -\e^{acb} & 0
                             \end{array}\right)
&{(J_c^d)_{\hat a}}^{\hat b}&=i\, \left(\begin{array}{cc}
                             \d_a^d \d_c^b-\d_a^b \d_c^d& 0\\
                              0 & -\d_b^d \d_c^a+\d_b^a \d_c^d
                             \end{array}\right) \,.
\end{align}
This decomposition provides a way to project out the spin connection and find the exact relation between the vielbein $\Omega_{\hat a}$ and the matrix $\Omega$
\begin{equation}
 \Omega_{\hat c}=\frac12 \Tr(E_{\hat c}\,\Omega)~.
\end{equation}
After having introduced all the necessary ingredients, we are ready to rewrite the Lagrangian in a form which resembles the $AdS_5\times S^5$ case. We separate it into
\begin{equation}
 L = L_B + L_F^{(2)} + L_F^{(4)}
\end{equation}
where the bosonic contribution is simply given by the standard bosonic sigma model with $AdS_4\times \mathbb{CP}^3$ as target space
\begin{equation}
L_B = \gamma^{ij}\left[\frac{e^{-4\varphi}}{4}\left(\partial_ix^+\partial_jx^-+
\partial_ix^1\partial_jx^1\right)+\partial_i\varphi
\partial_j\varphi+{\Omega^a}_i{\Omega_a}_j\right]
\end{equation}
where the vielbein ${\Omega^a}_i$ are defined in the natural way 
${\Omega^a}={\Omega^a}_i\, d\s^i$ with $\s^i=(\t,\s)$. 
Notice also that ${\Omega^{\hat a}}_i\,{\Omega_{\hat a}}_j=2\, {\Omega^a}_i\,{\Omega_a}_j$
 for the symmetry of the worldsheet metric. The quadratic part in the fermion fields can be expressed 
 as
\begin{align}\label{lagnew}
L_F^{(2)} = -2\, e^{-4\vf} \pa_i x^+ \Big[&\frac{i}{2} \g^{ij} \big(\eta^{\hat{a}}D_j\eta_{\hat{a}}+
\theta^{\hat{a}}D_j\theta_{\hat{a}}-2\, \Omega_j^{\hat c}\, \eta E_{\hat c}\eta \big)+\varepsilon^{ij} 
\eta^{\hat{a}} {C_{\hat a}}^{\hat{b}} \big(D_j \theta_{\hat b}+e^{-2\vf} \eta_{\hat b} \pa_j x^1) \nonumber\\
 +&\frac{i}{2} \g^{ij} \big(\bar\eta^{4}\pa_j\eta_{4}+\bar\theta^{4}\pa_j\theta_{4}
 -4\, i\, \eta_a\Omega^{a}_j\eta_4+ 2\, i\, \Omega_{a\phantom{a}j}^{\phantom{a}a} \Theta-h.c.\big)\nonumber \\
 +&\frac{1}{2}\varepsilon^{ij}\big(\bar \eta^4 \pa_j \theta_4-\bar \theta^4 \pa_j \eta_4 +4\, i\, \eta_a\Omega^{a}_j\theta_4 +2\,i\, \Omega_{a\phantom{a}j}^{\phantom{a}a} \tilde\Theta-e^{-2\vf} \Theta \pa_j x^1+h.c.\big)
 \Big] \,.
\end{align}
Here we have introduced the charge conjugation matrix $C$, given explicitly by\footnote{The fact that the matrix is diagonal and not anti-diagonal is a consequence of our conventions for grouping the spinors. Notice also that for our conventions $\eta^{\hat a}\eta_{\hat a}=0$ whereas $\eta^{\hat a} {C_{\hat a}}^{\hat b} \eta_{\hat b}=-2 \eta_a \bar\eta^a$.}
\begin{equation}
 {C_{\hat a}}^{\hat b}=\left(\begin{array}{cc}
                              \d_a^b & 0 \\
                              0 & -\d^a_b
                             \end{array}\right),
\end{equation}
and the combinations $\Theta=\theta_4\bar{\theta}^4+\eta_4\bar{\eta}^4$ 
and $\tilde \Theta =\theta_4\bar{\eta}^4-\eta_4\bar{\theta}^4$. 
The first line of this Lagrangian \eqref{lagnew} closely resembles expression (1.6) 
of \cite{Metsaev:2000yf}, that is the $AdS_5\times S^5$ Lagrangian in Wess-Zumino type parametrization. 
This is the part of the Lagrangian that does not contain the fermions $\eta_4$ and $\theta_4$, which emerge \cite{Uvarov:2009hf} when 
obtaining the \adscp action from dimensional reduction of the $AdS_4 \times S^7$ 
supermembrane action.
%This is in contrast with the $OSp(4|6)/\left( SO(1,3)\times U(3) \right)$ supercoset formulation, 
%where only 24 fermionic coordinates arise. However, in the latter setting the $\kappa$-symmetry 
%light-cone gauge fixing for strings moving in the $AdS$ space only (as for the light-like cusp) is not viable. 
%On the contrary the former approach \cite{Uvarov:2009hf} allows to impose a $\kappa$-symmetry 
%gauge condition that consists of setting to zero all the coordinates associated to fermionic generators 
%with negative charge with respect to the $SO(1,1)$ generator $M^{+-}$ (see \cite{Uvarov:2009hf} and \cite{Metsaev:2000yf} for the $AdS_5\times S^5$ case). 
The main difference with respect to $AdS_5\times S^5$ is that the $SU(4)$ R-symmetry 
is not explicitly realized on the fermionic Lagrangian \eqref{lagnew}. 
This feature is inherited by the quantum fluctuations around the light-like cusp. 
As a result of the broken symmetry, the spectrum contains fermionic degrees of freedom with 
different masses (one gets $6$ massive and $2$ massless excitations). 
Our one- and two-loop calculations have explicitly shown that the role of the 
massless fermions ($\tilde\eta_4$ and $\tilde\theta_4$) is crucial for compensating 
the bosonic degrees of freedom, making the one-loop partition function UV-finite. 
At two loops their interactions with the other excitations would in principle 
start playing a part. Nevertheless it turns out that the massless fermions 
decouple from the computation and do not contribute to the two-loop result.

The last term of the superstring Lagrangian is quartic in fermions
\begin{equation}
L_F^{(4)} = 4\, e^{-8\vf} \g^{ij} \pa_i x^+\pa_j x^+   [(\eta_a\bar{\eta}^a)^2+2\,\varepsilon^{abc}
\eta_a\eta_b\eta_c\eta_4+2
\eta_4\bar{\eta}^4\eta_a\bar{\eta}^a-\Theta^2+h.c.] \,.
\end{equation}
As discussed for the quadratic part, the first terms clearly reminds the expression for $AdS_5\times S^5$ (equation (1.10) of \cite{Metsaev:2000yf}), whereas the others contain the non-trivial interactions of $\eta_4$ and $\theta_4$.

%%%%%%%%%%%%%%%%%%%%%%%%%%%%%%%%%
\section{Details on the expanded Lagrangian}\label{expansion}
\label{app:lagr_exp}
%%%%%%%%%%%%%%%%%%%%%%%%%%%%%%%%%
In this appendix we provide the details of the Lagrangian \eqref{eq:Lagrangian_exp} expanded up to quartic order. 
As in section \ref{twoloops} we list the vertices as they appear  in ${\cal L}_{int}$, 
namely with an extra factor $\frac12$ with respect to the original Lagrangian.
We drop tildas, understanding that we are dealing with the fluctuation fields 
of \eqref{eq:Lagrangian_exp}. 
The cubic vertices are
\begin{equation*}
V_{\varphi x^1 x^1}=-4\,\varphi\, \left[(\pa_s-{\textstyle\frac12})\, x^1\right]^2 \qquad V_{\varphi^3}=2\varphi\left[(\pa_t \varphi)^2-(\pa_s\varphi)^2\right] \qquad V_{\varphi |z|^2}=2\varphi\left[|\pa_t z|^2-|\pa_s z|^2\right]
\end{equation*}
\begin{align}
V_{z\eta\eta}&=-\e^{abc} \pa_t \bar z_a \eta_b \eta_c + h.c. &
 V_{z\eta\theta}&=-2\, \e^{abc} \bar z_a \eta_b (\pa_s-{\textstyle\frac12})\theta_c-h.c. \nonumber\\
 V_{\varphi\eta \theta}&=-4\, i\, \varphi\, \eta_a (\pa_s-{\textstyle\frac12})\bar\theta^a -h.c. &
 V_{x^1 \eta \eta}&=-4\, i\, \bar \eta^a \eta_a (\pa_s-{\textstyle\frac12}) x^1 \nonumber\\
 V_{z\eta_a\eta_4}&=\ -2\, \pa_t z^a \eta_a \eta_4 + h.c. &
 V_{z\eta_a\theta_4}&=2\, \pa_s z^a \eta_a \theta_4-h.c. \nonumber\\
 V_{\varphi \eta_4 \bar\theta^4}&= -2\,i\, \varphi\, (\bar\theta^4 \pa_s \eta_4-\pa_s\bar\theta^4 \eta_4)-h.c. &
 V_{x^1 \bar\psi^4 \psi_4}&= -2\, i\, (\bar\eta^4 \eta_4+\bar\theta^4 \theta_4) (\pa_s-{\textstyle\frac12}) x^1 
\end{align}
The quartic vertices  read
\begin{align}
V_{z^4}&=\frac16 \left[ \left(\bar z_a \pa_t z^a\right)^2 + \left(\bar z_a \pa_s z^a\right)^2 + \left(z^a \pa_t \bar z_a\right)^2 + \left(z^a \pa_s \bar z_a\right)^2
\right. & &\nonumber\\& \hspace{1cm} \left.
- |z|^2\left(|\pa_t z|^2+|\pa_s z|^2\right) - \left|\bar z_a \pa_t z^a\right|^2 - \left|\bar z_a \pa_s z^a\right|^2\right]
\end{align}
\begin{align}
V_{\varphi^2 x^1 x^1}&=16\,\varphi^2\, \left[(\pa_s-{\textstyle\frac12})\, x^1\right]^2 &
V_{\varphi^4}&=4\, \varphi^2\left[(\pa_t \varphi)^2+(\pa_s\varphi)^2+\frac16\varphi^2\right]\\
V_{\varphi^2 |z|^2}&=4\, \varphi^2 \left[|\pa_t z|^2+|\pa_s z|^2\right]&
V_{\dot z\bar z \bar\psi^4 \psi_4}&= -2\, i\, (\bar\eta^4 \eta_4+\bar\theta^4 \theta_4)\bar z _b\pa_t z^b+h.c.\\
V_{\eta^2 \eta_4 \bar\eta_4}&= 8\,  \bar \eta^4 \eta_4 \bar \eta^a \eta_a  &
 V_{ z'\bar z \bar\psi^4 \psi_4}&= -2\, i\, (\bar\eta^4 \theta_4-\bar\theta^4 \eta_4)\bar z _b\pa_s z^b-h.c.\\
 V_{\eta^4}&=4 (\bar\eta^a \eta_a)^2&
 V_{\varphi^2 \eta_4 \bar\theta^4}&= 4\,i\, \varphi^2\, (\bar\theta^4 \pa_s \eta_4-\pa_s\bar\theta^4 \eta_4)-h.c.\\
  V_{\eta_4 \bar\eta_4\theta_4 \bar\theta_4}&=-8\, \bar \eta^4 \eta_4 \bar \theta^4 \theta_4&
  V_{\varphi\, x^1 \bar\psi^4 \psi_4}&=12\,i\, \varphi\, (\bar\eta^4 \eta_4+\bar\theta^4 \theta_4) (\pa_s-{\textstyle\frac12}) x^1\\
 V_{\eta^3 \eta_4}&=4\,  \e^{abc}\eta_a\eta_b\eta_c\eta_4+h.c.&
V_{zz\bar\eta^a\eta_4}&=-2\, i\, \e_{abc} \pa_t z^a z^b \bar \eta^c \eta_4+ h.c.\\
V_{\varphi\, z\eta_a\theta_4}&=-8\, \varphi\,  \pa_s z^a \eta_a \theta_4-h.c. &
 V_{\varphi\, z\eta\theta}&=8\, \varphi \e^{abc} \bar z_a \eta_b (\pa_s-{\textstyle\frac12})\theta_c-h.c. \\
 V_{zz\bar\eta^a\theta_4}&=2\, i\, \e_{abc} \pa_s z^a z^b \bar \eta^c \theta_4- h.c.&
  V_{zz\eta\eta}&=-2\, i\, (\bar z_a \pa_t z^a \bar\eta^b \eta_b-\bar z_b \pa_t z^a \bar \eta^b \eta_a) +h.c. \\
  V_{\varphi\, x^1 \eta \eta}&=24\, i\,\varphi\, \bar \eta^a \eta_a (\pa_s-{\textstyle\frac12}) x^1 &
  V_{zz\eta\theta}&=-2\, i\, [|z|^2 \eta_a (\pa_s-{\textstyle\frac12})\bar\theta^a - \bar z_b z^a\eta_a (\pa_s-{\textstyle\frac12})\bar\theta^b] -h.c. \\
  V_{\varphi^2 \eta \theta}&=8\, i\, \varphi^2\, \eta_a (\pa_s-{\textstyle\frac12})\bar\theta^a -h.c. &
 V_{x^1 z\eta\eta}&=-4\, (\pa_s-{\textstyle\frac12})x^1 \e^{abc} \bar z_a \eta_b \eta_c - h.c. 
\end{align} 
%%%%%%%%%%%%%%%%%%%%%%%%%%%%%%%%%
\section{Integral reductions}\label{app:intred}
In this appendix we provide the relevant tensor integral reductions in two dimensions that 
we used in the computation of the two-loop correction to the partition function.
We define the two basic scalar integrals
\begin{align}
I\left(m^2\right) & \equiv \int \frac{d^2p}{\left(2\pi\right)^2}\, \frac{1}{p^2+m^2} \\
I\left(m_1^2,m_2^2,m_3^2\right) & \equiv \int \frac{d^2p\, d^2q\,d^2r}{\left(2\pi\right)^4}\,  \frac{\delta^{(2)}(p+q+r)}{(p^2+m_1^2)(q^2+m_2^2)(r^2+m_3^2)} \,. 
\end{align}
Then we have (the factors $(2\pi)^4$ in the denominator of the integrands are understood)
{\allowdisplaybreaks
\begin{align}
& \int \frac{d^2p\, d^2q\,d^2r\, p^{\mu} q^{\nu}\, \delta^{(2)}(p+q+r)}{(p^2+m_1^2)(q^2+m_2^2)(r^2+m_3^2)} = \\& 
= \frac{\delta^{\mu\nu}}{4} \left[ I(m_1^2)I(m_2^2) - I(m_1^2)I(m_3^2) - I(m_2^2)I(m_3^2) + (m_1^2+m_2^2-m_3^2) I(m_1^2,m_2^2;m_3^2) \right]
\\&
I^{\mu}_{\mu}(m_1^2,m_2^2;m_3^2)=\int \frac{d^2p\, d^2q\,d^2r\, \left(p\cdot q\right)\, \delta^{(2)}(p+q+r)}{(p^2+m_1^2)(q^2+m_2^2)(r^2+m_3^2)} = \\& 
= \frac{1}{2} \left[ I(m_1^2)I(m_2^2) - I(m_1^2)I(m_3^2) - I(m_2^2)I(m_3^2) + (m_1^2+m_2^2-m_3^2) I(m_1^2,m_2^2;m_3^2) \right]
\\&
\int \frac{d^2p\, d^2q\,d^2r\, p^{\mu}\, p^{\nu}\, \delta^{(2)}(p+q+r)}{(p^2+m_1^2)(q^2+m_2^2)(r^2+m_3^2)} 
= \frac{\delta^{\mu\nu}}{2} \left[ I(m_2^2)I(m_3^2) - m_1^2\, I(m_1^2,m_2^2;m_3^2) \right]
\\&
J\equiv \int \frac{d^2p\, d^2q\,d^2r\, p^2 q^2\, \delta^{(2)}(p+q+r)}{(p^2+m_1^2)(q^2+m_2^2)(r^2+m_3^2)} 
= m_1^2 m_2^2\, I(m_1^2,m_2^2;m_3^2) - m_1^2\, I(m_1^2)I(m_3^2) - m_2^2\, I(m_2^2)I(m_3^2) 
\\&
K\equiv \int \frac{d^2p\, d^2q\,d^2r\, (p\cdot q)^2\, \delta^{(2)}(p+q+r)}{(p^2+m_1^2)(q^2+m_2^2)(r^2+m_3^2)} 
= \frac{1}{2} \left[ - m_2^2\, I(m_2^2)I(m_3^2) - m_1^2\, I(m_1^2)I(m_3^2) + \right.\nonumber\\&\left. \hspace{6.5cm} + (m_1^2+m_2^2-m_3^2) I^{\mu}_{\mu}(m_1^2,m_2^2;m_3^2) \right]
\\&
\int \frac{d^2p\, d^2q\,d^2r\, p^{\mu}\, p^{\nu}\, q^{\rho}\, q^{\sigma}\, \delta^{(2)}(p+q+r)}{(p^2+m_1^2)(q^2+m_2^2)(r^2+m_3^2)} 
= \left(\frac38 J - \frac14 K \right) \delta^{\mu\nu}\delta^{\rho\sigma} + 
\left(\frac14 K - \frac18 J \right) \left( \delta^{\mu\rho}\delta^{\nu\sigma} + \delta^{\mu\sigma}\delta^{\nu\rho} \right)
\\&
\int \frac{d^2p\, d^2q\,d^2r\, p^{\mu}\, p^{\nu}\, p^{\rho}\, q^{\sigma}\, \delta^{(2)}(p+q+r)}{(p^2+m_1^2)(q^2+m_2^2)(r^2+m_3^2)} 
= \frac18 \left( \delta^{\mu\nu}\delta^{\rho\sigma} + \delta^{\mu\rho}\delta^{\nu\sigma} + \delta^{\mu\sigma}\delta^{\nu\rho} \right)\nonumber\\& \hspace{6.cm}
\left[ m_2^2\, I(m_2^2)I(m_3^2) - m_1^2\, I^{\mu}_{\mu}(m_1^2,m_2^2;m_3^2) \right]
\\&
L\equiv \int \frac{d^2p\, d^2q\,d^2r\, p^2\, \left(q\cdot r\right)\, \delta^{(2)}(p+q+r)}{(p^2+m_1^2)(q^2+m_2^2)(r^2+m_3^2)} 
= -m_1^2\, I^{\mu}_{\mu}(m_3^2,m_2^2;m_1^2)
\\&
M\equiv \int \frac{d^2p\, d^2q\,d^2r\, (p\cdot q) (p\cdot r)\, \delta^{(2)}(p+q+r)}{(p^2+m_1^2)(q^2+m_2^2)(r^2+m_3^2)} 
= \frac{1}{2} \left[ (m_1^2+m_3^2-m_2^2) I^{\mu}_{\mu}(m_1^2,m_2^2;m_3^2) + 
\right.\nonumber\\&\left. \hspace{7.cm} + m_1^2\, I(m_1^2)I(m_3^2) - m_2^2\, I(m_2^2)I(m_3^2)  \right]
\\&
\int \frac{d^2p\, d^2q\,d^2r\, p^{\mu}\, p^{\nu}\, q^{\rho}\, r^{\sigma} \delta^{(2)}(p+q+r)}{(p^2+m_1^2)(q^2+m_2^2)(r^2+m_3^2)} 
= \left(\frac38 L - \frac14 M \right) \delta^{\mu\nu}\delta^{\rho\sigma} + 
\left(\frac14 M - \frac18 L \right) \left( \delta^{\mu\rho}\delta^{\nu\sigma} + \delta^{\mu\sigma}\delta^{\nu\rho} \right).
\end{align}}
%%%%%%%%%%%%%%%%%%%%%%%%%%%%%%%%%
%%%%%%%%%%%%%%%%%%%%%%%%%%%%%%%%%
\bibliographystyle{nb}
\bibliography{2loop}
%%%%%%%%%%%%%%%%%%%%%%%%%%%%%%%%%

%%%%%%%%%%%%%%%%%%%%%%%%%%%%%%%%%
\end{document}